\documentclass[12pt]{article}
\usepackage{setspace}
\usepackage{authblk}
\usepackage{amsthm,amsmath,amssymb}
\usepackage{cases}
\usepackage{dsfont}
\usepackage{multirow}
\usepackage{float}
\usepackage{dsfont}

\numberwithin{equation}{section}

\def\ep{\varepsilon}

\newtheorem{thm}{Theorem}[section]

\newtheorem{corollary}{Corollary}[section]

\newtheorem{remark}{Remark}[section]

\begin{document}

\date{}
\title{Inference from Small and Big Data Sets with Error Rates }

\author{Mikl\'{o}s Cs\"{o}rg\H{o}\thanks{mcsorgo@math.carleton.ca} and Masoud M. Nasari\thanks{mmnasari@math.carleton.ca}    \\
\small{School of Mathematics and Statistics of Carleton University}\\\small{ Ottawa, ON, Canada} }

\maketitle

\begin{abstract}
\noindent
In this paper we introduce  randomized $t$-type statistics that will be referred  to as \emph{randomized pivots}. We show that these randomized pivots yield central limit theorems with a significantly smaller magnitude of error as compared to that of their  classical counterparts under the same conditions. This constitutes a desirable result when a relatively small number of data is available.
When a data set is too big to be processed, we use our randomized pivots to make inference about the mean  based on  significantly smaller sub-samples. The approach taken is shown to relate naturally to  estimating distributions of both small and big data sets.
\end{abstract}

%\small{Keywords:
%}

% \linenumbers

\section{\normalsize{Introduction} }\label{section 1}
In this paper we address the  problem of making inference   about the population mean when  the available sample  is  either small or  big. In case of having a small sample we develop a randomization technique that yields   central limit theorems (CLT's) with a significantly smaller  magnitude of error that would compensate  for the lack of sufficient information as a result of having a small sample. Our technique works even when the sample is so small that  the classical CLT cannot be used to  make a valid inference.  In the case of having a big sample, we also develop a technique to make inference about the mean based on  a smaller sub-sample that can be  drawn  without dealing with the entire original data set that may not be even processable.

\par
Unless stated otherwise, $X, X_{1},\ldots$ throughout   are assumed to be independent random variables with a common distribution function $F$ (i.i.d. random variables), mean $\mu:=E_X X$ and variance $0<\sigma^{2}:=E_X  (X-\mu)^2< +\infty$.
Based on $X_1, \ldots,X_n$, a random sample on $X$,  for each integer  $n\geq 1$, define
\begin{equation*}
\bar{X}_{n}:=\sum_{i=1}^{n}X_i\big/n \ \textrm{and} \  S^{2}_{n}:=\sum_{i=1}^{n}(X_i -\bar{X}_{n})^2\big/ n,
\end{equation*}
the sample mean and sample variance, respectively, and consider the classical Student $t-$statistic
\begin{equation}\label{def. of T_n}
T_{n}(X):= \frac{\bar{X}_n}{S_n/\sqrt{n}}=\frac{\sum_{i=1}^n X_i}{S_n \sqrt{n}}
\end{equation}
that, in turn, on replacing $X_i$ by $X_i-\mu$, $1\leq i \leq n$, yields
\begin{equation}\label{equivalent to T_n}
T_{n}(X-\mu):= \frac{\bar{X}_n -\mu}{S_n/\sqrt{n}}=\frac{\sum_{i=1}^n (X_i-\mu)}{S_n \sqrt{n}},
\end{equation}
the classical Student $t$-pivot for the population mean $\mu$.
\par
Define now $T_{m_n,n}^{(1)}$ and  $G_{m_n,n}^{(1)}$, randomized versions of $T_{n}(X)$ and $T_{n}(X-\mu)$ respectively, as follows:

\begin{equation}\label{T^{*}}
T^{(1)}_{m_n,n}:= \frac{\bar{X}_{m_n,n}-\bar{X}_n   }{S_{n}\sqrt{\sum_{i=1}^{n} (\frac{w^{(n)}_i}{m_n}-\frac{1}{n})^{2}}  }=\frac{\sum_{i=1}^{n} \big( \frac{w^{(n)}_{i}}{m_{n}} -\frac{1}{n}   \big)   X_{i}    }{S_{n}\sqrt{\sum_{i=1}^{n} (\frac{w^{(n)}_i}{m_n}-\frac{1}{n})^{2}}  },
\end{equation}

\begin{equation}\label{G^{*}}
G^{(1)}_{m_n,n}:= \frac{\sum_{i=1}^{n} \big| \frac{w^{(n)}_{i}}{m_{n}} -\frac{1}{n}   \big| \big( X_{i}-\mu \big)   }{S_{n}\sqrt{\sum_{i=1}^{n} (\frac{w^{(n)}_i}{m_n}-\frac{1}{n})^{2}}  },
\end{equation}
where,
\begin{equation}\label{bootstrap mean}
\bar{X}_{m_n,n}:=\sum_{i=1}^n w^{(n)}_i X_i/m_{n},
\end{equation}
is the randomized sample mean and  the weights $(w^{(n)}_{1},\ldots,w^{(n)}_{n})$ have a multinomial distribution of size
$m_n:= \sum_{i=1}^n  w^{(n)}_{i}$ with respective probabilities  $1/n$, i.e.,
\begin{equation*}
  (w^{(n)}_{1},\ldots,w^{(n)}_{n})\  \substack{d\\=}\ \ multinomial(m_{n};\frac{1}{n},\ldots,\frac{1}{n}).
\end{equation*}
\par
The just introduced respective randomized $T^{(1)}_{m_n,n}$ and  $G^{(1)}_{m_n,n}$ versions of $T_{n}(X)$ and $T_{n}(X-\mu)$ can be computed via  re-sampling from  the set of indices $\{ 1,\ldots,n\}$ of   $X_1,\ldots,X_n$ with replacement $m_n$ times so  that, for each $1\leq i \leq n$,  $w^{(n)}_i$ is  the count of the  number of times the index $i$ of $X_i$ is chosen in this  re-sampling process.

\begin{remark}\label{Remark 1}
In view of the preceding definition of $w^{(n)}_{i}$, $1\leq i \leq n$, they form a row-wise independent triangular array of random variables such that  $\sum_{i=1}^n  w^{(n)}_{i}=m_n$ and, for each $n\geq 1$,
$$(w^{(n)}_{1},\ldots,w^{(n)}_{n})\  \substack{d\\=}\ \ multinomial(m_{n};\frac{1}{n},\ldots,\frac{1}{n}),
$$
i.e., the weights  have a multinomial distribution of size $m_n$ with respective probabilities $1/n$.
Clearly,  for each $n$, $w^{(n)}_{i}$    are independent from the random sample $X_{i}$, $1\leq i \leq n$.
Weights denoted by $w_{i}^{(n)}$ will stand for triangular multinomial random variables in this context throughout.
\end{remark}
Thus, $T^{(1)}_{m_n,n}$ and $G^{(1)}_{m_n,n}$ can simply be computed by generating, independently from the data, a realization of the random multinomial weights $(w_{1}^{(n)},\ldots,w_{n}^{(n)})$ as in Remark \ref{Remark 1}.

\par
Define the similarly computable further randomized versions $T^{(2)}_{m_n,n}$ and $G^{(2)}_{m_n,n}$ of $T_{n}(X)$ and $T_{n}(X-\mu)$  respectively, as follows:

\begin{eqnarray}
T^{(2)}_{m_n,n}&:=& \frac{\bar{X}_{m_n,n}-\bar{X}_n   }{S_{m_n,n}\sqrt{\sum_{i=1}^{n} (\frac{w^{(n)}_i}{m_n}-\frac{1}{n})^{2}}  }=\frac{\sum_{j=1}^{n} \big( \frac{w^{(n)}_{i}}{m_{n}} -\frac{1}{n}   \big)   X_{i}   }{S_{m_n,n}\sqrt{\sum_{i=1}^{n} (\frac{w^{(n)}_i}{m_n}-\frac{1}{n})^{2}}  } \label{T^{**}}\\
G^{(2)}_{m_n,n}&:=& \frac{\sum_{i=1}^{n} \big| \frac{w^{(n)}_{i}}{m_{n}} -\frac{1}{n}   \big| \big( X_{i}-\mu \big)   }{S_{m_n,n}\sqrt{\sum_{i=1}^{n} (\frac{w^{(n)}_i}{m_n}-\frac{1}{n})^{2}}  }, \label{G^{**}}
\end{eqnarray}
where $S_{m_n,n}^2$ is the randomized sample variance,  defined as
\begin{equation}\label{def. of S^*}
S^{2}_{m_n,n}:= \sum_{i=1}^n w_{i}^{(n)} \big( X_i- \bar{X}_{m_n,n} \big)^2 \big/ m_n.
\end{equation}

 \par
Unlike $T_{n}(X)$  that can be transformed into $T_{n}(X-\mu)$, the Student pivot for $\mu$ as in (\ref{equivalent to T_n}) (cf. Gin\'{e} \emph{et al}. \cite{Gine Gotze Mason} for the asymptotic equivalence of the two), its randomized  versions $T_{m_n,n}^{(1)}$ and $T_{m_n,n}^{(2)}$ do not have this straightforward  property, i.e., they do  not yield a pivotal quantity for the population mean $\mu=E_{X} X$ by simply replacing each  $X_i$ by $X_i-\mu$ in their  definitions. We introduced $G_{m_n,n}^{(1)}$ and $G_{m_n,n}^{(2)}$ in this paper to serve as direct randomized pivots for the population mean $\mu$, while $T_{m_n,n}^{(1)}$ and $T_{m_n,n}^{(2)}$ will now be viewed on their own as randomized pivots for the  sample mean $\bar{X}_n$ in case of a big data set.

\par
Our Theorem \ref{Berry-Esseen} and its corollaries will explain the higher order accuracy these randomized pivots provide for inference about the mean $\mu$, as compared to that provided by $T_{n}(X-\mu)$.

\par
Among the many outstanding contributions in the literature studying the asymptotic behavior of $T_{n}(X)$ and $T_{n}(X-\mu)$,  our main tool  in this paper,  Theorem \ref{Berry-Esseen} below, relates mostly  to       Bentkus \emph{et al}. \cite{Bentkus et al}, Bentkus and G\"{o}tze \cite{Bentkus and Gotze},  Pinelis \cite{Pinelis} and   Shao \cite{Shao}.

\par
A short outline of the contributions of this paper reads as follows.
\par
In Section \ref{berry-esseen section} we derive the rates  of convergence for $G^{(i)}_{m_n,n}$ and $T^{(i)}_{m_n,n}$, $i=1,2$, via establishing   Berry-Ess\'{e}en type results in Theorem \ref{Berry-Esseen} and its Corollaries \ref{The Rate}-\ref{The joint Rate n}. In Corollary \ref{The joint Rate n} we show that, on taking $m_n=n$, $G^{(i)}_{m_n,n}$ and $T^{(i)}_{m_n,n}$, $i=1,2$, converge, in distribution,  to the standard normal at the rate of $O(1/n)$. This rate is significantly better than the best possible $O(1/\sqrt{n})$ rate of convergence under  similar moment conditions   for the classical $t$-statistic $T_{n}(X)$ and its Student pivot $T_{n}(X-\mu)$, based on a random sample of size $n$.
 The latter $O(1/\sqrt{n})$ rate is best possible in the sense that it cannot be improved  without restricting the class of distribution functions  of the data,   for example, to normal or  symmetrical distributions. In section \ref{berry-esseen section} we also present  numerical studies  that well  support our conclusion that, on taking $m_n=n$, $G^{(i)}_{m_n,n}$ and $T^{(i)}_{m_n,n}$, $i=1,2$, converge to standard normal at a significantly faster rate than that of  the classical CLT.
In Sections \ref{section pivot for mu} and \ref{section pivot for bar{X}}, the respective rates of convergence of the CLT's in Section \ref{berry-esseen section} will be put to  significant use.
In Section \ref{section pivot for mu},  $G^{(i)}_{m_n,n}$, $i=1,2$, are studied as natural  asymptotic pivots for the population mean $\mu=E_{X} X$. In section \ref{section pivot for bar{X}},  $T^{(i)}_{m_n,n}$, $i=1,2$, are studied as natural asymptotic pivots for the sample mean $\bar{X}_n$ that  closely  shadows   $\mu$, when dealing with big data sets of univariate observations of $n$ labeled units $\{X_1,\ldots,X_n \}$.
In this case, instead of trying to process the entire data  set that may even be impossible to do, sampling it indirectly via generating random weights independently from the data as in Remark \ref{Remark 1} makes it possible to use  $T^{(2)}_{m_n,n}$  to construct an interval estimation for the sample mean  $\bar{X}_n$  based on   significantly smaller sub-samples which can be obtained without dealing directly  with the entire data set (cf. Remark \ref{Remark 4.2'}).  The latter confidence set for $\bar{X}_n$ in turn will be seen to contain the population mean $\mu$ as well, and with same rates of convergence, in terms $m_n$ and $n$, as those established for having $\bar{X}_n$ in there.
 In Section \ref{CLT empirical} the sample and population distribution functions are studied along the lines of  Sections \ref{berry-esseen section}-\ref{section pivot for bar{X}}. The proofs are given in Sections \ref{Proofs} and Appendices 1 and 2.

\par
For   throughout use, we let $(\Omega_X,\mathfrak{F}_X,P_X)$ denote  the probability space of the random variables  $X,X_1,\ldots$, and $(\Omega_w,\mathfrak{F}_w,P_w)$ be  the probability space on which the weights $$\big(w^{(1)}_1,(w^{(2)}_1,w^{(2)}_{2}),\ldots,(w^{(n)}_1,\ldots,w^{(n)}_{n}),\ldots \big)$$
 are defined. In view of the independence of these two sets of random variables, jointly they live on the direct product probability space $(\Omega_X \times \Omega_w, \mathfrak{F}_X \otimes \mathfrak{F}_w, P_{X,w}=P_X .\ P_w )$. For each $n\geq 1$, we also let  $P_{.|w}(.)$   stand for the conditional probabilities  given $\mathfrak{F}^{(n)}_w:=\sigma(w^{(n)}_1,\ldots,w^{(n)}_{n})$  with corresponding conditional expected value $E_{.|w}(.)$.

\section{\normalsize{The rate of convergence of the CLT's for   $G_{m_n}^{(i)}$ and $T_{m_n}^{(i)}$, $i=1,2$ }}\label{berry-esseen section}
One of the  efficient  tools  to control  the error when   approximating the distribution function  of a statistic with that  of a standard normal random variable  is provided by   Berry-Ess\'{e}en type inequalities  (cf., e.g.,  Serfling \cite{Serfling}),
  which provide   upper bounds  for the error of approximation  for any finite number of observations
in hand.
It is well known that, on assuming $E_{X} |X-\mu|^3<+\infty$,  as  the sample size  $n$ increases to infinity,   the rate at which the Berry-Ess\'{e}en    upper bound  for $\sup_{-\infty< t <+\infty} |P_{X}(T_{n}(X-\mu)\leq t)-\Phi(t)|$  vanishes is $O(n^{-1/2})$, where, and also throughout,   $\Phi$ stands for the standard normal distribution function.

Furthermore, the latter  rate is best possible in the sense that it cannot be improved without narrowing the class of distribution functions considered.

\par
Our Berry-Ess\'{e}en type inequalities   for the respective conditional, given the weights $w^{(n)}_i$'s, distributions of  $G^{(1)}_{m_n,n}$ and $T_{m_n,n}^{(1)}$,  as in   (\ref{G^{*}}) and (\ref{T^{*}}) respectively, and $G^{(2)}_{m_n,n}$ and $T_{m_n,n}^{(2)}$,  as in  (\ref{G^{**}}) and (\ref{T^{**}}) respectively, read as follows.

\begin{thm}\label{Berry-Esseen}
Assume that $E_{X}|X|^{3}<+\infty$ and let $\Phi(.)$ be the standard normal  distribution function.
Also, for arbitrary positive numbers $\delta,\varepsilon$, let $\varepsilon_1, \varepsilon_2>0$ be so that $\delta> (\varepsilon_1/\varepsilon)^2+P_{X}(|S^{2}_n -\sigma^2|>\varepsilon^{2}_1 )+\varepsilon_2 >0$, where, for $t\in \mathbb{R}$, $\Phi(t-\varepsilon)-\Phi(t)>- \varepsilon_2$ and $\Phi(t+\varepsilon)-\Phi(t)<  \varepsilon_2$.
 Then, for all $n,m_n$    we have

\begin{eqnarray*}
(A)&&P_{w}\big\{ \sup_{-\infty < t < +\infty}\Big| P_{X|w} (G^{(1)}_{m_n,n}\leq t)-\Phi(t)  \Big|>\delta \big\}~~~~~~~~~~~~~~~~~~~~~~~~~~~~\\
&&\leq \delta^{-2}_n(1-\varepsilon)^{-3}(1-\frac{1}{n})^{-3 } (\frac{n}{m^{3}_n}+\frac{n^{2}}{m^{3}_n})\{ \frac{15 m^{3}_n}{n^3}+ \frac{25 m^{2}_n}{n^2}+\frac{m_n}{n} \}  \\
&&+ \varepsilon^{-2}\frac{m^{2}_n}{ (1-\frac{1}{n})} \Big\{ \frac{1-\frac{1}{n}}{n^3 m^{3}_n } + \frac{(1-\frac{1}{n})^4}{m^{3}_n}  + \frac{(m_n -1)(1-\frac{1}{n})^2}{n m^{3}_n} +  \frac{4(n-1)}{n^3 m_n}       +\frac{1}{m^{2}_n} \\
&&- \frac{1}{n m^{2}_n} + \frac{n-1}{n^{3} m^{3}_{n}} + \frac{4(n-1)}{n^2 m^{3}_{n}} - \frac{(1-\frac{1}{n})^2}{m^{2}_n}        \Big\},
%+ \varepsilon P_{X} \big( \big| S_{n}^{2} -\sigma^{2}  \big| > \varepsilon\big)
\end{eqnarray*}
and also
\begin{eqnarray*}
(B)&&P_{w}\big\{ \sup_{-\infty < t < +\infty}\Big| P_{X|w} (T_{m_n,n}^{(1)}\leq t)-\Phi(t)  \Big|>\varepsilon \big\}~~~~~~~~~~~~~~~~~~~~~~~~~~~~\\
&&\leq \delta^{-2}_n(1-\varepsilon)^{-3}(1-\frac{1}{n})^{-3 } (\frac{n}{m^{3}_n}+\frac{n^{2}}{m^{3}_n})\{ \frac{15 m^{3}_n}{n^3}+ \frac{25 m^{2}_n}{n^2}+\frac{m_n}{n} \}  \\
&&+ \varepsilon^{-2}\frac{m^{2}_n}{ (1-\frac{1}{n})} \Big\{ \frac{1-\frac{1}{n}}{n^3 m^{3}_n } + \frac{(1-\frac{1}{n})^4}{m^{3}_n}  + \frac{(m_n -1)(1-\frac{1}{n})^2}{n m^{3}_n} +  \frac{4(n-1)}{n^3 m_n}       +\frac{1}{m^{2}_n} \\
&&- \frac{1}{n m^{2}_n} + \frac{n-1}{n^{3} m^{3}_{n}} + \frac{4(n-1)}{n^2 m^{3}_{n}} - \frac{(1-\frac{1}{n})^2}{m^{2}_n}        \Big\},
%+ \varepsilon P_{X} \big( \big| S_{n}^{2} -\sigma^{2}  \big|> \varepsilon \big),
\end{eqnarray*}
where
\begin{equation*}
\delta_n:= \frac{\delta- (\varepsilon_1/\varepsilon)^2 -P_{X} ( |S^{2}_n -\sigma^2|>\varepsilon^{2}_1 )+\varepsilon_2   }{C E_{X}|X-\mu|^{3}/\sigma^{3/2}},
\end{equation*}
with $C$ being  a universal constant as in  the Berry-Ess\'{e}en upper bound for independent and not necessarily identically distributed summands (cf. page 33 of Serfling \cite{Serfling}).
\end{thm}

\par
The following result, a corollary to Theorem \ref{Berry-Esseen}, gives the rate of convergence of the respective conditional  CLT's for $G^{(1)}_{m_{n},n}$ and $T_{m_n,n}^{(1)}$, as well as for $G^{(2)}_{m_{n},n}$ and $T_{m_n,n}^{(2)}$.

\begin{corollary}\label{The Rate}
Assume that $E_{X}|X|^3<+\infty$. If $n, \ m_n \rightarrow +\infty$ in such a way that $m_n=o(n^2)$,  then, for  arbitrary $\delta>0$,  we have
\begin{equation*}
(A) \ P_{w}\big\{ \sup_{-\infty < t < +\infty}\Big| P_{X|w} (G^{(1)}_{m_n,n}\leq t)-\Phi(t)  \Big|>\delta \big\} = O\Big( \max\{ \frac{m_n}{n^2}, \frac{1}{m_n} \}  \Big), ~~~~~~~~~~~~~
\end{equation*}
\begin{equation*}
(B) \ P_{w}\big\{ \sup_{-\infty < t < +\infty}\Big| P_{X|w} (T_{m_n,n}^{(1)}\leq t)-\Phi(t)  \Big|>\delta \big\} = O\Big( \max\{ \frac{m_n}{n^2}, \frac{1}{m_n}  \}  \Big). ~~~~~~~~~~~~~
\end{equation*}
Moreover, if $E_{X} X^4 <+\infty$, if $n,m_n \to +\infty$ in such a way that $m_n=o(n^2)$ and $n=o(m^{2}_n)$ then, for  $\delta>0$,  we also have
\begin{equation*}
(C) \ P_{w}\big\{ \sup_{-\infty < t < +\infty}\Big| P_{X|w} (G^{(2)}_{m_n,n}\leq t)-\Phi(t)  \Big|>\delta \big\} = O\Big( \max\{ \frac{m_n}{n^2}, \frac{1}{m_n},\frac{n}{m^{2}_{n}} \}  \Big), ~~~~~~~~~~~~~
\end{equation*}
\begin{equation*}
(D) \ P_{w}\big\{ \sup_{-\infty < t < +\infty}\Big| P_{X|w} (T^{(2)}_{m_n,n}\leq t)-\Phi(t)  \Big|>\delta \big\} = O\Big( \max\{ \frac{m_n}{n^2}, \frac{1}{m_n},\frac{n}{m^{2}_{n}} \}  \Big). ~~~~~~~~~~~~~
\end{equation*}
\end{corollary}

\par
When $0< E_{X} X^2 <+\infty$, the conditional $P_{X|w}$ CLT's for $G^{(i)}_{m_n,n}$ and $T^{(i)}_{m_n,n}$, $i=1,2$, whose respective rates of convergence are established in Corollary \ref{The Rate},  can be concluded  as direct consequences of  a realization of the  Lindeberg-Feller CLT (cf. Theorems 27.3 and 27.4 of Billingsley \cite{Billinsley}) as formulated in Lemma 5.1 of Cs\"{o}rg\H{o} \emph{et al.} \cite{Scand} (cf. also Appendix 2) that is also known as the H\'{a}jeck -Sid\'{a}k Theorem (cf., e.g., Theorem 5.3 in DasGupta \cite{Das Gupta}).

\begin{remark}\label{Remark 2.1}
On taking $m_n=n$, when $E_{X}|X|^{3}<+\infty$, the rates of convergence of Corollary \ref{The Rate} for both $G^{(1)}_{m_n,n}$ and $T_{m_n,n}^{(1)}$ are of order
$O( n^{-1})$.  The same is true for $G^{(2)}_{m_n,n}$ and $T^{(2)}_{m_n}$ for $m_n=n$ when $E_{X} X^{4}<+\infty$.
\end{remark}

\begin{remark}\label{Remark 2.2}
When $E_{X} X^4 <+\infty$, the extra term   $n/m_{n}^{2}$ which appears in the rate of convergence of   $G^{(2)}_{m_n,n}$ and $T^{(2)}_{m_n,n}$ in (C) and (D) of Corollary \ref{The Rate}, is the rate at which $P_{w}\big\{ P_{X|w}\big( |S^{2}_{m_n,n}-S^{2}_{n}|>\varepsilon_1\big) > \varepsilon_2 \big\}$ approaches zero as $n,m_n \to +\infty$,  where  $\varepsilon_1$ and $\varepsilon_2$ are arbitrary positive numbers.
\end{remark}

\par
The conditional CLT's resulting from (A), (B), (C) and (D) of Corollary \ref{The Rate}  imply respective unconditional CLT's in terms of the joint distribution $P_{X,w}$ as in the following Corollaries \ref{The joint Rate m_n} and \ref{The joint Rate n}.

\begin{corollary}\label{The joint Rate m_n}
Assume that $E_{X}|X|^{3}<+\infty$. If $n, \ m_n \rightarrow +\infty$ in such a way that $m_n=o(n^2)$,  then, for  arbitrary $\delta>0$,  we have

\begin{eqnarray}
\sup_{-\infty < t < +\infty}  \big|  P_{X,w} (G^{(1)}_{m_n,n}\leq t)-\Phi(t) \big| &\leq& \delta+ O\Big( \max\{ \frac{m_n}{n^2}, \frac{1}{m_n} \}  \Big), \label{joint dis1}\\
\sup_{-\infty < t < +\infty}\Big| P_{X,w} (T_{m_n,n}^{(1)}\leq t)-\Phi(t)  \Big| &\leq& \delta+ O\Big( \max\{ \frac{m_n}{n^2}, \frac{1}{m_n} \}  \Big). \label{joint dis2}
\end{eqnarray}
Moreover, if $E_{X} X^4 <+\infty$, if $n,m_n \to +\infty$ in such a way that $m_n=o(n^2)$ and $n=o(m^{2}_n)$ then, for arbitrary $\delta>0$,  we also have
\begin{eqnarray}
\sup_{-\infty < t < +\infty}  \big|  P_{X,w} (G^{(2)}_{m_n,n}\leq t)-\Phi(t) \big| &\leq& \delta+ O\Big( \max\{ \frac{m_n}{n^2}, \frac{1}{m_n},\frac{n}{m^{2}_{n}} \}  \Big) \label{joint dis33}\\
\sup_{-\infty < t < +\infty}  \big|  P_{X,w} (T^{(2)}_{m_n,n}\leq t)-\Phi(t) \big| &\leq& \delta+ O\Big( \max\{ \frac{m_n}{n^2}, \frac{1}{m_n},\frac{n}{m^{2}_{n}} \}  \Big) \label{joint dis44}
\end{eqnarray}

\end{corollary}

The following Corollary \ref{The joint Rate n},  a trivial consequence of Corollary \ref{The joint Rate m_n} on taking $m_n=n$, is   of particular interest as it asserts  that the rate at which each of the error terms  of the CLT's therein vanishes happens at the optimal  $O(1/n)$ rate.  This  is a significant improvement over the classical Berry-Ess\'{e}en   $O(1/\sqrt{n})$ rate of error for $T_{n}(X)$ and $T_{n}(X-\mu)$  on only assuming the same $E|X|^3 <+\infty$ moment condition for $G^{(1)}_{n,n}$ and $T^{(1)}_{n,n}$, and $E_{X} X^4 <+\infty$ as well in the case of $G^{(2)}_{n,n}$ and $T^{(2)}_{n,n}$. Further moment conditions would not improve the $O(1/n)$ rates of convergence in hand, as below.

\begin{corollary}\label{The joint Rate n}
When $m_n=n$,   for arbitrary positive $\delta$,  as $n\rightarrow +\infty$, we have

\begin{eqnarray}
\sup_{-\infty < t < +\infty}  \big|  P_{X,w} (G^{(1)}_{n,n} \leq t)-\Phi(t) \big| &\leq& \delta+ O(1/n), \label{joint dis11}\\
\sup_{-\infty < t < +\infty}\Big| P_{X,w} (T^{(1)}_{n,n}\leq t)-\Phi(t)  \Big| &\leq& \delta+ O(1/n), \label{joint dis22}\\
\sup_{-\infty < t < +\infty}  \big|  P_{X,w} (G^{(2)}_{n,n}\leq t)-\Phi(t) \big| &\leq& \delta+ O(1/n), \label{joint dis3}\\
\sup_{-\infty < t < +\infty}  \big|  P_{X,w} (T^{(2)}_{n,n}\leq t)-\Phi(t) \big| &\leq& \delta+ O(1/n), \label{joint dis4}
\end{eqnarray}
where (\ref{joint dis11}) and (\ref{joint dis22}) hold true when $E_{X} |X|^3 <+\infty$, and (\ref{joint dis3}) and (\ref{joint dis4}) hold true when $E_{X} X^4 <+\infty$.

\end{corollary}

\begin{remark}\label{Remark 2.4}
The respective conclusions of Corollaries \ref{The joint Rate m_n} and \ref{The joint Rate n} amount to saying that the indicated upper bounds become arbitrary small at the indicated $O(.)$ rates when $n,m_n \rightarrow +\infty$ as postulated for each, under their respective moment conditions. As an illustration of what we mean by this, we spell out  statement (\ref{joint dis11}) accordingly. Thus, on assuming $E_{X}|X|^3 < +\infty$, fore any given arbitrary small $ \varepsilon>0$, take the arbitrary $\delta>0$ in (\ref{joint dis11}) to be so  small that $\varepsilon-\delta>0$. Consequently, there exists an $n_0=n(\varepsilon,\delta)$ so that for $n\geq n_0$, $O(1/n)<\varepsilon-\delta$, and thus, for all $n\geq n_0$, $\delta+O(1/n)<\varepsilon$, i.e., the indicated upper bound in (\ref{joint dis11}) becomes arbitrary small at the rate $O(1/n)$  as $n \rightarrow +\infty$.        \end{remark}

\section{Numerical Studies}
In this section we use the statistical software R to conduct our numerical  studies for comparing the performance of $G^{(1)}_{n,n}$ as in (\ref{joint dis11}) of Corollary \ref{The joint Rate n} to that of its classical counterpart $T_{n}(X-\mu)$.
\par
In order to provide initial motivation for the  more  in-depth numerical studies as in  Tables 2 and 3 below, that indicate  a significantly better performance  of the pivot  $G^{(1)}_{n,n}$ for $\mu$   over its classical counterpart $T_{n}(X-\mu)$, we first  compare  the empirical probabilities of  coverage   of these pivots for $\mu$ in Table 1. The nominal probability coverage for the one sided confidence intervals (C.I.'s)  in Table 1 is $95\%$ in terms of the standard normal cutoff point 1.644854. The C.I.'s in Table 1 are based on      1000 replications of the data $(X_1, \ldots, X_n)$ for both pivots $G^{(1)}_{n,n}$ and $T_n (X-\mu)$,  and 1000 replication of $(w^{(n)}_1, \ldots, w^{(n)}_n)$, with $\sum_{i=1}^n w^{(n)}_i=n$, for computing $G^{(1)}_{n,n}$.   The intervals are   obtained by setting:

\begin{equation}\nonumber
G^{(1)}_{n,n} \leq 1.644854 ~~~~~ \textrm{and} ~~~~~~~~ T_{n} (X-\mu ) \leq 1.644854.
\end{equation}
The empirical probabilities of  coverage for  each one of these pivots are presented in  Table 1 for the distributions therein.

\begin{table}[ht]\label{Table 1}
\caption{\small{Comparing the empirical probability coverage of   pivot $G^{(1)}_{n,n}$ to $T_{n}(X-\mu)$} }
\vspace{-.2 cm}
\begin{center}
\begin{tabular}{ c|c|c|c  }
\hline \hline
  Distribution\ of \ Sample&  $n$ & $\textrm{coverage \ of} \ G^{(1)}_{n,n}$ & $\textrm{coverage \ of} \ T_n(X-\mu)$  \\
\hline
\multirow{2}{*}{Binomial$(10,0.1)$} &20& 0.956 &0.964 \\
                                    &30&  0.953 &0.960 \\
                                    %&40& 0.634& 0.466 \\
                         \hline

 \multirow{2}{*}{Exponential$(1)$}   &20&  0.959 & 0.975  \\
                                    &30&   0.956  & 0.968  \\
                                   % &40&  0.432 & 0.044   \\

\hline
\multirow{2}{*}{Normal$(0,1)$}      &20&  0.945 & 0.931  \\
                                    &30&  0.951 & 0.946  \\
                                    %&40&  0.634  & 0.612   \\
                                    \hline
\multirow{2}{*}{Beta$(5,1)$}        &20&  0.914 & 0.903  \\
                                    &30&  0.949 & 0.909  \\
                                    %&40&  0.634  & 0.612   \\
                                    \hline
\multirow{2}{*}{Binomial$(10,.9)$}      &20&  0.922 &0.904  \\
                                        &30&  0.956 & 0.936  \\
                                        %&40&  0.634  & 0.612   \\
                                    \hline
\end{tabular}
\end{center}
\end{table}
Table 1 below shows that the sampling distribution of  $G^{(1)}_{n,n}$ in each case, even for small sample sizes,   is close enough  to the standard  normal distribution. Using standard normal  percentiles,   $G^{(1)}_{n,n}$, as a pivot for the population mean $\mu$, tends to yield    probabilities of  coverage that are near to the nominal $95\%$ even for  sample sizes for which  the classical CLT for $T_{n}(X-\mu)$  fails to provide  valid C.I.'s for $\mu$.

\newpage
In order to study  in-depth   the refinement  provided by $G^{(1)}_{n,n}$ over the classical $T_n (X-\mu)$ in view of (\ref{joint dis11}) of Corollary \ref{The joint Rate n}, in the following Tables 2 and 3 we   present some numerical illustrations of the  rates of convergence of    \emph{one sided} C.I.'s for the population mean $\mu$ based on the pivot $G^{(1)}_{n,n}$ whose validity  and the rate at which they approach to their nominal probability coverage   are concluded  in
(\ref{joint dis11}) of our   Corollary \ref{The joint Rate n} for $G^{(1)}_{n,n}$. In Table 2 the empirical probability coverage of these asymptotic C.I.'s  based on the pivot $G^{(1)}_{n,n}$ with nominal  $95\%$ level are compared to the empirical probability coverage of the  \emph{exact}  size $t$-C.I.'s based on the  pivot       $T_n(X-\mu)$ whose exact sampling distribution is Student-$t$ with $n-1$ degrees of freedom   when the data are i.i.d. normal.
\par
To construct our \emph{asymptotic} $95\%$   C.I.'s based on $G^{(1)}_{n,n}$ in both Tables 2 and 3,  we use the standard normal $95\%$ cutoff point $1.644854$. In Table 2  we use  exact cutoff points of the Student $t$-statistic $T_n(X-\mu)$, valid for   \emph{exact} C.I.'s for the population mean. All of the one sided C.I.'s in Table 3 are asymptotic, with both  pivots in hand having  standard  normal limiting distribution as $n\rightarrow +\infty$.

\par
 Tables 2 and 3 display the proportion of  500 generated one sided   C.I.'s with empirical coverage probability value in $[0.94,0.96]$. Each one of these  500 C.I.'s is constructed by generating 500   sets  of i.i.d. observations $(X_1,\ldots,X_n)$, with $n$ as displayed,  from the  indicated respective underlying distributions.  For  simulating  each value of  $G^{(1)}_{n,n}$, we also  generate 500 sets of the multinoimal weights $(w_{1}^{(n)},\ldots,w_{n}^{(n)})$, with $\sum_{1\leq i \leq n}w_{i}^{(n)}=n$ and associated probability vector  $(1/n,\ldots,1/n)$.

\par
Both  Tables 2 and 3 indicate a highly satisfactory performance of the pivot $G^{(1)}_{n,n}$ even when it is compared to an exact size Student $t$-confidence interval as in Table 2.

\par
To exhibit   the  performance of the pivot $G^{(1)}_{n,n}$  in Table 3, in addition to normal, we also consider data from skewed  distributions.  It is known that the Student $t$-distribution converges to standard normal at a rate of order  $O(1/n)$. The numerical results in Table 3 show that, based on normal data,  $G^{(1)}_{n,n}$  performs as good as  the $t$-statistic $T_{n}(X-\mu)$. The latter is an empirical  indication that $G^{(1)}_{n,n}$ converges to standard normal at the rate of  $O(1/n)$.

\par
In both Tables 2 and 3, we denote the  proportions of the C.I.'s with empirical probability coverage values between $94\%$ and $96\%$  associated with the pivots $G^{(1)}_{n,n}$ and $T_{n}(X-\mu)$, respectively,  by $prop \ G^{(1)}$ and $prop \ T_{n}(X-\mu)$.

\begin{table}[ht]\label{Table 2}
\caption{\small{Comparing the  pivot $G^{(1)}_{n,n}$ to the Student  $t$-distribution} }
\vspace{-.2 cm}
\begin{center}
\begin{tabular}{ c|c|c|c  }
\hline \hline
  Distribution of Sample &  $n$ & $\textrm{prop}\ G^{(1)}$ & $\textrm{prop}\ T_n(X-\mu)$  \\
\hline
  \multirow{3}{*}{Normal$(0,1)$} &20& 0.55 &0.626 \\
                                 &25& 0.622 &0.662   \\
                                 &30& 0.628 &0.632 \\
\hline
\end{tabular}
\end{center}
\end{table}
\noindent
In Table 2  the standard norma 95\% cutoff point 1.644854 is used  for the pivot $G_{n,n}^{(1)}$
and  the cutoff points $t_{0.05,19}=1.729$, $t_{0.05,24}=1.711$ and $t_{0.05,29}=1.699$ were used for the pivot $T_{n}(X-\mu)$ for $n=20$, $n=25$ and $n=30$, respectively.\\
%$\vspace {-0.4 cm}$
\begin{table}[ht]\label{Table 3}
\caption{\small{Comparing the pivot $G^{(1)}_{n,n}$ to $T_{n}(X-\mu)$} }
\vspace{-.2 cm}
\begin{center}
\begin{tabular}{ c|c|c|c  }
\hline \hline
  Distribution\ of \ Sample&  $n$ & $\textrm{prop}\ G^{(1)}$ & $\textrm{prop}\ T_n(X-\mu)$  \\
\hline
\multirow{3}{*}{Binomial$(10,0.1)$} &20& 0.745 & 0.486 \\
                                    &30&  0.764& 0.546 \\
                                    &40&  0.768& 0.511 \\
                         \hline

  \multirow{3}{*}{Poisson$(1)$} &20& 0.552 &0.322 \\
                                &30& 0.554 &0.376   \\
                                &40& 0.560 &0.364 \\
                         \hline

 \multirow{3}{*}{Lognormal$(0,1)$}     &20& 0.142& 0.000  \\
                                       &30& 0.168& 0.000   \\
                                       &40& 0.196& 0.000 \\
                         \hline

 \multirow{3}{*}{Exponential$(1)$}   &20&  0.308 & 0.016  \\
                                    &30&  0.338 & 0.020  \\
                                    &40&  0.432 & 0.044   \\

\hline
\multirow{3}{*}{Normal$(0,1)$}      &20&  0.566  & 0.486  \\
                                    &30&  0.600 & 0.568  \\
                                    &40&  0.634  & 0.612   \\
                                    \hline
\multirow{3}{*}{Beta$(5,1)$}        &20&  0.074 & 0.000   \\
                                    &30&  0.136 & 0.016 \\
                                    &40&  0.234 & 0.058   \\
                                    \hline
\end{tabular}
\end{center}
\end{table}
In Table 3 the standard normal 95\% cutoff point 1.644854 was used for both pivots $G^{(1)}_{n,n}$  and $T_{n}(X-\mu)$. Furthermore, in Table 3 Lognormal(0,1) stands for a Lognormal distribution with mean  zero and variance one.

\newpage

\section{\normalsize{Randomized asymptotic pivots for the population mean $\mu$}}\label{section pivot for mu}
We are now to present $G_{m_n,n}^{(1)}$ of (\ref{G^{*}}) and $G_{m_n,n}^{(2)}$  of (\ref{G^{**}}) as \emph{direct} asymptotic randomized  pivots for the population mean $\mu=E_{X} X$, first  when only $0<\sigma^2:=E_{X}(X-\mu)^2<+\infty$ is assumed, followed by assuming $E_{X} |X|^3 <+\infty$ as in Remark \ref{Remark 3.1}, and $E_{X} X^4 <+\infty$ as in Remark \ref{Remark 3.2}.

\par
We note that for the coinciding  numerator terms  of $G_{m_n,n}^{(1)}$ and $G_{m_n,n}^{(2)}$ we have
\begin{equation}\label{(1.22)new}
E_{X|w}\big( \sum_{i=1}^n |\frac{w_{i}^{(n)}}{m_n}-\frac{1}{n}|(X_i-\mu) \big)=0.
\end{equation}
Furthermore, given $w^{(n)}_i$'s, for the randomized  weighted average
\begin{equation}\label{(1.21)}
{\sum_{i=1}^n |\frac{w_{i}^{(n)}}{m_n}-\frac{1}{n}| (X_i-\mu)}=: \bar{X}_{m_n,n}(\mu),%{\sum_{j=1}^n |\frac{w_{j}^{(n)}}{m_n}-\frac{1}{n}|}
\end{equation}
%is an unbiased estimator of the population mean $\mu$. Moreover,
mutatis mutandis in verifying (\ref{(5.30)}) in Appendix 1, we conclude that  when the original sample size $n$ \emph{is fixed} and $m:=m_n$, then,  \emph{as} $m \to +\infty$, we have %$\sum_{i=1}^n |\frac{w_{i}^{(n)}}{m_n}-\frac{1}{n}|(X_i-\mu) \to 0$ in $probability-P_{X,w}$, as well as
\begin{equation}\label{(1.22)}
\bar{X}_{m_n,n}(\mu)=\bar{X}_{m,n}(\mu) \to 0 \ in \ probability-P_{X,w},
\end{equation}
and the same holds true if $n\to +\infty$ as well.
\par
In view of (\ref{(1.22)new})
\begin{equation}\label{(1.22)'}
\frac{\sum_{i=1}^n |\frac{w_{i}^{(n)}}{m_n}-\frac{1}{n}| X_i}{\sum_{j=1}^n |\frac{w_{j}^{(n)}}{m_n}-\frac{1}{n}|}=: \hat{X}_{m_n,n}.
\end{equation}
is an unbiased estimator for $\mu$ with respect to $P_{X|w}$.

\par
It can be shown that when $E_X X^2<+\infty$, as  $n,m_n \to +\infty$ such that $m_n=o(n^2)$, $\hat{X}_{m_n,n}$ is a consistent estimator for the population mean $\mu$ in terms of $P_{X,w}$, i.e.,
\begin{equation}\label{doubleprime}
\hat{X}_{m_n,n} \to \mu \ in \ probability-P_{X,w}.
\end{equation}

\par
In Appendix 1 we give a  direct proof for  (\ref{doubleprime})  for the important case when $m_n=n$, for which the  CLT's in Corollary  \ref{The Rate} hold true  at the  $O(1/n)$ rate.

\par
As to $G_{m_n,n}^{(1)}$ of (\ref{G^{*}}), on replacing $\big( \frac{w_{i}^{(n)}}{m_n}-\frac{1}{n}   \big)$  by $\big| \frac{w_{i}^{(n)}}{m_n}-\frac{1}{n} \big|$ in the proof of (a) of Corollary 2.1 of Cs\"{o}rg\H{o} \emph{et al.} \cite{Scand} (cf. Appendix 2), \emph{as $n,m_n \to +\infty$ so that} $m_n=o(n^2)$, when $0< \sigma^2:= E_{X} (X-\mu)^2 <0$, we arrive at
\begin{equation}\label{(1.23)}
P_{X|w} (G_{m_n,n}^{(1)} \leq t) \to \Phi(t) \ in \ probability-P_w\ for \ all \ t\in \mathds{R},
\end{equation}
and, via Lemma 1.2 in S. Cs\"{o}rg\H{o} and Rosalsky \cite{Csorgo and Rosalsky},   we conclude also the unconditional CLT
\begin{equation}\label{(1.24)}
P_{X,w} (G_{m_n,n}^{(1)} \leq t) \to \Phi(t) \ for \ all \ t\in \mathds{R}.
\end{equation}

\begin{remark}\label{Remark 3.1}
When $E_{X} |X|^3 <+\infty$ and $n,m_n \to +\infty$ so that $m_n=o(n^2)$, then, in addition to (\ref{(1.23)}), we have (A) of Corollary  \ref{The Rate} as well, and, in addition to (\ref{(1.24)}), we also have  (\ref{joint dis1})   and (\ref{joint dis11}) as in Corollaries \ref{The joint Rate m_n} and \ref{The joint Rate n} respectively.
\end{remark}

\par
When $E_{X} X^2 <+\infty$, in   Appendix 1 we   show that when $n$ \emph{is fixed} and $m:=m_n \rightarrow +\infty$, the randomized sample variance $S^{2}_{m_n,n}$,  as defined in (\ref{def. of S^*}),
converges in probability-$P_{X,w}$ to the sample variance $S_{n}^2$, i.e.,   we have (cf. (\ref{(5.31)}) in Appendix 1 or Remark 2.1 of
Cs\"{o}rg\H{o} \emph{et al}. \cite{Scand})
\begin{equation}\label{retaining S^2 with fixed n}
S^{2}_{m,n} \rightarrow S_{n}^{2} \ in \ probability-P_{X,w}.
\end{equation}

\par
For related results along these lines in terms of $u$- and $v$-statistics,   we refer to Cs\"{o}rg\H{o} and Nasari \cite{J. multivariate Anal}, where, in a more general setup, we establish in probability and almost sure consistencies  of randomized  $u$- and $v$-statistics.
\par
In   Appendix 1 we also show that, when $E_X X^2<+\infty$,  \emph{if} $n,\ m_n\rightarrow +\infty$ \emph{so that} $n=o(m_n)$,   then we have (cf. (\ref{(5.31)}) in Appendix 1)
\begin{equation}\label{bias of S*}
\big( S^{2}_{m_n,n} - S_{n}^{2}\big) \rightarrow 0 \ in \ probability-P_{X,w}.
\end{equation}

\par
When $E_{X} X^4<+\infty$, the preceding convergence also holds true when $n=o(m^{2}_n)$ (cf. the proof of (C) and (D) of Corollary \ref{The Rate}).
 \par
 On combining (\ref{bias of S*}) with the CLT in (\ref{(1.24)}), when $E_X X^2<+\infty$, \emph{as} $n,m_n \to +\infty$ \emph{so that $m_n=o(n^2)$ and} $n=o(m_n)$, the following unconditional CLT holds true   as well in terms of $P_{X,w}$
\begin{equation}\label{(1.25)}
G_{m_n,n}^{(2)} \substack{d\\ \longrightarrow}\ Z ,
\end{equation}
where, and also throughout, ${\substack{d\\ \longrightarrow}}$ stands for convergence in distribution,    $G_{m_n,n}^{(2)}$ is as defined in (\ref{G^{**}}), and $Z$ stands for a standard normal random variable.

\begin{remark}\label{Remark 3.2}
Assuming that $E_{X} X^4 <+\infty$ and $n,m_n \to +\infty$ so that $m_n=o(n^2)$ and $n=o(m^{2}_n)$, then we have  (\ref{joint dis33}) and (\ref{joint dis3}) as in Corollaries \ref{The joint Rate m_n} and \ref{The joint Rate n} respectively, i.e.,  then the unconditional CLT
\begin{equation}\label{(3.10)}
G^{(2)}_{m_n,n}\ {\substack {d\\ \longrightarrow}} Z
\end{equation}
holds true in terms of $P_{X,w}$ at the therein indicated respective rates of convergence, and  we have (C) of Corollary \ref{The Rate} as well, i.e.,

\begin{equation}\label{(3.11)}
P_{X|w} ( G^{(2)}_{m_n,n}\leq t) \rightarrow \Phi(t) \ in \ probability-P_w\ for \ all \ t\in   \mathds{R}
\end{equation}
at the therein indicated rate of convergence.
\end{remark}

\par
With $G^{(1)}_{m_n,n}$ and $G^{(2)}_{m_n,n}$ in mind as direct asymptotic pivots for $\mu$, the CLT's as in (\ref{(1.23)}) and  (\ref{(1.24)}), as well as their respective versions  as spelled out in Remark \ref{Remark 3.1}, together with the  CLT's as in
(\ref{(1.25)}), (\ref{(3.10)})  and (\ref{(3.11)}), can be used to construct exact size asymptotic C.I.'s  for the population mean $\mu=E_{X} X$.  Thus,  in terms of $G^{(1)}_{m_n,n}$,  \emph{as} $n,m_n \rightarrow +\infty$ \emph{and} $m_n=o(n^2)$,   we conclude as follows,  a $1-\alpha$  size asymptotic C.I. for the population mean $\mu=E_{X} X$, which is  valid both in  terms of the conditional  $P_{X|w}$ and in unconditional $P_{X,w}$ distributions as in (\ref{(1.23)}) and (\ref{(1.24)}) respectively, as well as  with rates of convergence as in Remark \ref{Remark 3.1}:

\begin{equation}\label{(1.27)}
  \hat{X}_{m_n,n}-z_{\alpha/2}\frac{S_n \sqrt{.}}{\sum_{j=1}^n |\frac{w_{j}^{(n)}}{m_n}-\frac{1}{n}|}
\leq \mu \leq
 \hat{X}_{m_n,n}+z_{\alpha/2}\frac{S_n \sqrt{.}}{\sum_{j=1}^n |\frac{w_{j}^{(n)}}{m_n}-\frac{1}{n}|}
\end{equation}
where  $z_{\alpha/2}$ satisfies $P(Z \geq z_{\alpha/2})=\alpha/2$  and  $\sqrt{.}:= \sqrt{\sum_{j=1}^n (\frac{w_{j}^{(n)}}{m_n}-\frac{1}{n})^2}$.

\par
When  $E_{X} X^4<+\infty$,  then we can replace $S_n$ by $S_{m_n,n}$ in (\ref{(1.27)}), and then   the thus obtained $1-\alpha$ size asymptotic C.I. for the population mean $\mu$ holds true in terms of  ${G}_{m_n,n}^{(2)}$ via both of the respective CLT's as  in (\ref{(3.10)}) and (\ref{(3.11)}) with respective  rates of convergence as indicated in Remark \ref{Remark 3.2}.

\par
In view of Remark \ref{Remark 3.1}, on taking $m_n=n$, when $E_{X} |X|^3 <+\infty$, then both CLT's as in (\ref{(1.23)})  and (\ref{(1.24)}) hold true with a $O(1/n)$ rate of convergence (cf. Remark \ref{Remark 2.1} and (\ref{joint dis1}) of Corollary \ref{The joint Rate n}). Hence, the $1-\alpha$ size asymptotic C.I. for $\mu$ as in (\ref{(1.27)}) is also achieved at that rate in both cases. The same conclusion remains true on replacing $S_n$ by $S_{m_n,n }$  in (\ref{(1.27)}) and taking $m_n=n$ when $E_{X} X^4 <+\infty$ (cf. Remarks \ref{Remark 3.2} and  \ref{Remark 2.1}, and (\ref{joint dis3}) of Corollary \ref{The joint Rate n}).

\section{\normalsize{Randomized asymptotic pivots for the sample  and population means  of big data sets}}\label{section pivot for bar{X}}
The numerical characteristics of a given big data set should be  fairly close to their population counterparts.  For instance, the sample mean of a give data set  $\{ X_1,\ldots,X_n\}$ of large size $n$ will be seen  to  deviate from the population mean  only by a negligible error in the context of this paper. The same will be seen to be  true for the sample percentiles and their   population counterparts in Section \ref{CLT empirical}.

\par
When processing the entire big data set is not an option, then its numerical characteristics become  unobservable,  and hence unknown. Thus the    estimators of the unknown parameters   themselves are to be estimated as well.

\par
In this section we   construct confidence sets  for the sample mean, $\bar{X}_n$, of a large i.i.d. sample,  shadowing that of the population $\mu$. These confidence sets  can in turn  be used to serve as C.I.'s for  the population mean $\mu$, due to closeness of the two parameters in hand (cf.  (\ref{(1.20)}) and (\ref{(4.9)'})).

\par
To begin with, we consider the associated numerator term of $T_{m_n,n}^{(i)}$, $i=1,2$, and write
\\
\begin{eqnarray}
\sum_{i=1}^{n} (\frac{w^{(n)}_i}{m_n}-\frac{1}{n}) X_i &=& \frac{1}{m_n} \sum_{i=1}^{n} w^{(n)}_{i} X_i -\frac{1}{n}\sum_{i=1}^{n} X_i \label{bias}\\
&=& \bar{X}_{m_n,n}-\bar{X}_{n}. \nonumber
\end{eqnarray}
We note that when the original sample size $n$ \emph{is} assumed to be \emph{fixed}, then on taking \emph{only one} large  sub-sample of size $m:=m_n$,  via re-sampling the set of indices of the observations with replacement as in Remark \ref{Remark 1}, as $m \rightarrow +\infty$,  we have
\begin{equation}\label{retaning bar{X} with fixed n}
\bar{X}_{m,n} \rightarrow  \bar{X}_n \ in \ probability-P_{X,w}
\end{equation}
(cf. (\ref{(5.30)}) of Appendix 1).

\par
Further to (\ref{retaning bar{X} with fixed n}), \emph{as} $n,\ m_n\rightarrow +\infty$, then (cf. (\ref{(5.30)}) in Appendix 1)
\begin{equation}\label{bias of bar{X*}}
\big(\bar{X}_{m_n,n}- \bar{X}_n  \big) \rightarrow 0\ in\ probability-P_{X,w}.
\end{equation}

\par
As to $T_{m_n,n}^{(1)}$, and further to (\ref{bias of bar{X*}}), we have that $E_{X|w}\big( \sum_{i=1}^n (\frac{w_{i}^{(n)}}{m_n}-\frac{1}{n}) X_i \big)=0$ and, \emph{if} $n,m_n \rightarrow +\infty$ \emph{so that} $m_n=o(n^2)$, then (cf. part (a) of Corollary 2.1 of Cs\"{o}rg\H{o} \emph{et al}. \cite{Scand} and Appendix 2)
\begin{equation}\label{(1.11)}
P_{X|w} (T_{m_n,n}^{(1)}\leq t) \rightarrow P(Z\leq t)\ in \ probability-P_w\ for \ all \ t\in \mathds{R}.
\end{equation}
Consequently, as $n,m_n \to +\infty$ \emph{so that} $m_n=o(n^2)$, we arrive at
\begin{equation}\label{(1.12)}
P_{X,w} (T_{m_n,n}^{(1)}\leq t) \rightarrow P(Z\leq t)\ for \ all \ t\in \mathds{R},
\end{equation}
an unconditional CLT.

\begin{remark}\label{Remark 4.1}
When $E_{X} |X|^3<+\infty$ and $n,m_n \rightarrow +\infty$ so that $m_n=o(n^2)$, then, in addition to (\ref{(1.11)}), we have (B) of Corollary \ref{The Rate} as well, and in addition to (\ref{(1.12)}), we also have (\ref{joint dis2}) and (\ref{joint dis22}) as in Corollaries  \ref{The joint Rate m_n}  and \ref{The joint Rate n} respectively.
\end{remark}

\par
Furthermore, in view of the latter CLT and (\ref{bias of S*}), \emph{as} $n,m_n \rightarrow +\infty$ \emph{so that} $m_n=o(n^2)$ \emph{and}  $n=o(m_n)$, \emph{in terms of probability}-$P_{X,w}$ we conclude the unconditional CLT
\begin{equation}\label{(1.13)}
{T}_{m_n,n}^{(2)} {\substack{d\\ \longrightarrow}}\ Z,
\end{equation}
where $T_{m_n,n}^{(2)}$  is as defined in (\ref{T^{**}}).

\begin{remark}\label{Remark 4.2}
Assuming that $E_X X^4 <+\infty$ and $n,m_n \rightarrow +\infty$ so that $m_n=o(n^2)$ and $n=o(m^{2}_n)$, we then  have (\ref{joint dis44}) and (\ref{joint dis4}) as in Corollaries  \ref{The joint Rate m_n} and \ref{The joint Rate n} respectively, i.e., then  the unconditional CLT as in $(\ref{(1.13)})$,  in terms of $P_{X,w}$,  holds true at the therein indicated respective rates of convergence. Naturally, under the same conditions, as $n,m_n \rightarrow +\infty$,  we have (D) of Corollary \ref{The Rate} as well, i.e.,

\begin{equation}\label{(4.8)}
P_{X|w} ( {T}_{m_n,n}^{(2)} \leq t) \longrightarrow \Phi(t)\ in \ probability-P_{w} \ for \ all\ t\in \mathds{R}
\end{equation}
at the therein indicated rate of convergence.
\end{remark}

\begin{remark}\label{Remark 4.2'}
Considering that  our approach to randomizing the original sample  in this section   coincides with drawing a smaller sub-sample of size $m_n$ with replacement from the  original big data set $\{ X_1, \ldots,X_n\}$  via re-sampling its index set $\{1,\ldots,n \}$ as in Remark \ref{Remark 1}, it is important to note that in order to compute both $\bar{X}_{m_n,n}$ and $S^{2}_{m_{n},n}$, as in  (\ref{bootstrap mean}) and (\ref{def. of S^*}), respectively, only those $X_i$'s are needed whose $w^{(n)}_i \neq 0$. This means that both $\bar{X}_{m_n,n}$ and $S^{2}_{m_{n},n}$ are computable based only on the  smaller  sub-sample rather than the entire original big data set.
\end{remark}

\par
Under their respective conditions the CLT's as in  (\ref{(1.13)}) and (\ref{(4.8)}) can be used to construct confidence sets for the sample mean $\bar{X}_n$ that is an unknown parameter in our present context.

\par
We spell out the one based on $T_{m_n,n}^{(2)}$ as in (\ref{(1.13)}) that is also valid  in terms of (\ref{(4.8)}), i.e., both in the context of Remark \ref{Remark 4.2}.   Accordingly,  \emph{when $E_{X} X^4 <+\infty$ and $m_n,n \to +\infty$ so that $m_n=o(n^2)$} and $n=o(m_{n}^2)$, then for any $\alpha \in (0,1)$, we conclude  a $1-\alpha$ size asymptotic confidence set for $\bar{X}_n$,  at the indicated rates of convergence, as follows

\begin{equation}\label{(1.20)}
 \bar{X}_{m_n,n}-z_{\alpha/2} S_{m_n,n} \sqrt{.} \leq  \bar{X}_n  \leq \bar{X}_{m_n,n}+z_{\alpha/2} S_{m_n,n} \sqrt{.}  ,
\end{equation}
where $z_{\alpha/2}$ is as in (\ref{(1.27)}), and $\sqrt{.}:=\sqrt{\sum_{j=1}^n  (\frac{w_{i}^{(n)}}{m_n}- \frac{1}{n} )^2}$.

\par
 When $E_{X}|X|<+\infty$, as $n\rightarrow +\infty$, we have that $  \bar{X}_n -\mu =: \varepsilon_n=o(1)$, almost surely in $P_X$-probability, as $n\rightarrow +\infty$.  Since,   the original sample size $n$ of a big data set is already  very large to begin with,  $\varepsilon_n$ is already negligible with high $P_X$-probability. Consequently,  the confidence set (\ref{(1.20)}) for $\bar{X}_n$ can actually  be viewed as  a $(1-\alpha)$ size asymptotic C.I. as well  for the population mean $\mu$, by simply rewriting it as follows

\begin{equation}\label{(4.9)'}
\bar{X}_{m_n,n}-z_{\alpha/2} S_{m_n,n} \sqrt{.}\leq \mu +\varepsilon_n \leq \bar{X}_{m_n,n}+z_{\alpha/2} S_{m_n,n} \sqrt{.},
\end{equation}
where $z_{\alpha/2}$ and $\sqrt{.}$ are as in (\ref{(1.20)}).

\par
We emphasize that (\ref{(1.20)}) and (\ref{(4.9)'}) are identical statements under the conditions as spelled out right above (\ref{(1.20)}). The asymptotic negligibility of the  error sequence $\varepsilon_n$ in (\ref{(4.9)'}) can,  however, be studied on its own as $n \rightarrow +\infty$, freely from the identical conditions  for (\ref{(1.20)}) and (\ref{(4.9)'}) that $m_n=o(n^2)$ and $n=o(m^{2}_{n})$, as $n,m_n\rightarrow +\infty$.

\par
To further elaborate on the fact that (\ref{(4.9)'}) should work well as an asymptotic  $(1-\alpha)$ size C.I. for the population mean $\mu$ in the case of a big data set, we make  use of some well known classical results on the complete convergence of $\bar{X}_n$ to $\mu$ under two or more moment  conditions for $X$.

\par
We first mention the  Erd\H{o}s-Hsu-Robbins  theorem (cf. \cite{Erdos-Hsu-Robinson 1}, \cite{Erdos-Hsu-Robinson 2} and \cite{Erdos-Hsu-Robinson 3})  that concludes

\begin{equation}\nonumber
\sum_{n=1}^{+\infty} P_{X}  \big( |\bar{X}_n -\mu| >\epsilon \big)<+\infty,\ for \ every\ \epsilon>0,
\end{equation}
if and only if $E_{X} X^2<+\infty$. Thus, in addition to concluding that $\bar{X}_n -\mu=\varepsilon_n =o(1)$ almost surely-$P_X$, we also infer  that,  for any $\epsilon>0$, $\{P_{X} (| \varepsilon_n|>\epsilon) \}_{n=1}^{+\infty}$  approaches  zero at a rate faster than  $O(1/n)$. In other words, as $n\rightarrow +\infty$,  $\varepsilon_n$ approaches zero in probability-$P_X$ at a rate faster than the best possible rate of convergence for $T^{(2)}_{m_n,n}$ (cf. Corollary \ref{The joint Rate n}).
Therefore, even when assuming only a two moment condition,   (\ref{(4.9)'})  captures $\bar{X}_n$ and $\mu$ simultaneously with a high $P_X$-probability, that is, typically, $1-1/(n\log^{2} n)$.

\par
Further along these lines, we also mention the    Baum and Katz theorem \cite{Baum and Katz} that asserts
\begin{equation*}
\sum_{n=1}^ {+\infty} n^{r/p -2} P_{X} \big(  |\bar{X}_n -\mu|> \epsilon\ n^{1/p - 1}\big)< +\infty,
\end{equation*}
for every $\epsilon>0$ and some $p\in (0,2)$, if and only if $E_{X}|X|^r <+\infty$. Thus, when  $E_X X^4<+\infty$, then for a big sample of size $n=10^6$, for example, with $p=1$

\begin{equation}\nonumber
P_{X} ( | \varepsilon_n | \leq  \epsilon) \ {\substack{>\\ \approx} } \ 1-\frac{1}{ 10^{18} (\log 10^6)^2   }\ for \ any\ \epsilon>0.
\end{equation}
This shows that $\varepsilon_n=\bar{X}_n -\mu$ in (\ref{(4.9)'}) becomes arbitrarily  small at a very fast rate in probability-$P_X$ in terms of the  original big sample size $n$, without paying attention to how $n$ and $m_n$ relate to each other when arriving at the asymptotic $(1-\alpha)$ size confidence set for covering $\bar{X}_n$ as in (\ref{(1.20)}). Hence, the confidence set (\ref{(1.20)}) for the unknown sample mean $\bar{X}_n$ of a big data set of size $n$, viewed as in (\ref{(4.9)'}), is also seen to be an asymptotic $(1-\alpha)$ C.I. for the unknown population mean  $\mu$ under the same conditions that are used to arrive at having (\ref{(1.20)}).

\par
We now also  illustrate how one goes about constructing the coinciding random  boundaries in (\ref{(1.20)}) and (\ref{(4.9)'}) in general, and then in  case of having a big sample of size $n=10^6$, as a convenient example.

\par
First of all we  emphasize  that in the asymptotic confidence set  (\ref{(1.20)}) for $\bar{X}_n$ of a big data set, the bounds in hand,   are computed by generating, independently from the entire data set, a realization of the random multinomial weights $(w^{(n)}_{1},\ldots,w^{(n)}_n)$ as in Remark \ref{Remark 1}. Thus, instead of trying to process the entire big data set $\{X_1,\ldots,X_n \}$ in order to compute $\bar{X}_n$, sampling it only via its index set $\{ 1,\ldots,n\}$ as above, we end up estimating $\bar{X}_n$ in terms of a confidence set as in $(\ref{(1.20)})$ that  can be based on significantly smaller sub-samples of size $m_n$ of the entire big data set  of size $n$, without having to deal with the latter directly, whenever $E_X X^4 <+\infty$ and $m_n=o(n^2)$ and $n=o(m_{n}^2)$ (cf. Remark \ref{Remark 4.2}). In this case the rate of convergence of  the conditional CLT as in  (\ref{(4.8)}), as well as its unconditional CLT as in (\ref{(1.13)}), is

\begin{equation}\label{(4.10)}
O\big(\max\{\frac{m_n}{n^2}, \frac{1}{m_n}, \frac{n}{m^{2}_n} \}\big)
\end{equation}
in view of (D) of Corollary \ref{The Rate} and (\ref{joint dis44}) of Corollary \ref{The joint Rate m_n} respectively.

\par
We note that, on account of having $n=o(m_{n}^2)$, as $m_n,n\rightarrow +\infty$  we cannot consider taking $m_n=n^{1/2}$ in the context of (\ref{(4.10)}). We may however consider taking

\begin{equation}\label{(4.11)}
m_n=n^{1/2} n^{\delta},\ 0< \delta<1/2,
\end{equation}
and then the rate of convergence in (\ref{(4.10)}) reduces to

\begin{equation}\label{(4.12)}
O(n^{- 2 \delta}),\ 0< \delta <1/2.
\end{equation}

\par
For  example, on taking $\delta =1/4$, then $m_n= n^{3/4}$, and the rate of convergence for covering $\bar{X}_n$ as in $(\ref{(1.20)})$  becomes $O(n^{-1/2})$, that coincides with that of the classical CLT for the Student $t$-statistic and pivot (cf. (\ref{def. of T_n}) and (\ref{equivalent to T_n})). For instance, in this case, for a big sample of size $n=10^{6}$,  the CLT  of (\ref{(4.8)}) and its unconditional version for $T^{(2)}_{m_n,n}$  are both  applied with a sub-sample  of size $m_{10^{6}}= \sum_{i=1}^{10^6} w^{(10^6)}_{i}= (10^6)^{3/4}\approx 31,623$, where the random multinomially distributed weights $(w^{(10^6)}_1,\ldots,w^{(10^6)}_n)$ are generated independently from the data $\{ X_1,\ldots,X_{10^6}\}$ with respective probabilities $1/10^6$, i.e.,

\begin{equation}\label{(4.13)}
(w^{(10^6)}_1,\ldots,w^{(10^6)}_{10^6})\ {\substack{d\\=}}  \ multinomial (31,623; \ \frac{1}{10^6},\ldots,\frac{1}{10^6} ).
\end{equation}
These multinomial weights, in turn,  are used to construct a $(1-\alpha)$ size confidence set \emph{\`{a} la} (\ref{(1.20)}), covering the unobserved mean $\bar{X}_{10^6}$, as well as the unknown population mean $\mu$,  with an error proportional to $0.001$ (cf. (\ref{(4.12)}) with $\delta=1/4$).

\par
More reduction of the sub-sample size $m_n$ can, for example, be achieved by taking

\begin{equation}\label{(4.14)}
m_n= n^{1/2} \log\log n
\end{equation}
instead of  that in  (\ref{(4.11)}) and,  via (\ref{(4.10)}), arriving at the rate of convergence

\begin{equation}\label{(4.15)}
O(1/(\log \log n)^2)
\end{equation}
for the CLT's in hand, instead of that in (\ref{(4.12)}). For instance, if  we again consider having  a   big sample of size $n=10^6$, then (\ref{(4.14)}) yields    a sub-sample of size $m_n= 10^3 \log\log 10^6 \approx 2, 626$, and constructing a $(1-\alpha)$ size   confidence set  \emph{\`{a} la} (\ref{(1.20)}), will  cover the unobserved $\bar{X}_{10^6}$, as well as the unknown population mean $\mu$, with an error proportional to $1/(\log\log 10^6)^2 \approx 1/7$. The latter increased error, as compared to the previous example with respective sub-sample size  $m_{10^6}=31,623$, is due to the much reduced sub-sample of size $m_{10^6}= 2, 626$ in this context. This scenario can also be viewed in terms of using normal $z_{\alpha/2}$ percentiles for the Student $t$-pivot $T_{n}(X-\mu)$ when  estimating the population mean $\mu$ on the basis of $n=49$ i.i.d. observations with an error proportional to $1/\sqrt{49}=1/7$.

\section{\normalsize{Randomized CLT's and C.I.'s for the empirical and theoretical distributions with application to big data sets}}\label{CLT empirical}
Let $X,X_1,X_2, \ldots$ be independent real valued random variables with a common distribution function $F$ as before, but now  without assuming the existence of any finite moments for $X$. Let  $\{ X_1,\ldots,X_n\}$ be a random sample  of size $n\geq 1$ on $X$ and, for each $n$,  define the  empirical distribution function

\begin{equation}\label{(5.1)}
F_{n}(x):= \sum_{i=1}^n \mathds{1}(X_i \leq x)/n, \ x\in \mathbb{R},
\end{equation}
and the sample variance of the indicator variables $\mathds{1}(X_i \leq x)$
\begin{equation}\label{(5.2)}
S^{2}_{n}(x):=\frac{1}{n} \sum_{i=1}^n \big( \mathds{1}(X_i \leq x) - F_{n}(x)\big)^2= F_n (x)(1-F_n (x)),\ x\in\mathbb{R}.
\end{equation}

\par
With  $m_n=\sum_{i=1}^{n} w^{(n)}_{i} $ and   the multinomial weights as in Remark \ref{Remark 1},
\begin{equation*}
\big(w^{(n)}_{1},\ldots,w^{(n)}_{n}\big) {\substack{d\\=}} \ multinomial \big(m_n;\frac{1}{n},\ldots,\frac{1}{n}\big),
\end{equation*}
that are independent from the random sample  of  $n$ labeled units $\{X_1,\ldots,X_n \}$,
define the randomized standardized   empirical process
\begin{eqnarray}
\alpha^{(1)}_{m_{n},n}(x)&:=& \frac{\sum_{i=1}^n (\frac{w^{(n)}_{i}}{m_n}-\frac{1}{n}  ) \mathds{1}(X_i \leq x)}{\sqrt{F(x)(1-F(x))} \sqrt{\sum_{j=1}^n (\frac{w^{(n)}_{j}}{m_n}-\frac{1}{n}  )^2} } \label{(5.3)}\\
&=& \frac{\sum_{i=1}^n \frac{w^{(n)}_{i}}{m_n} \mathds{1}(X_i \leq x)-F_{n}(x)}{\sqrt{F(x)(1-F(x))} \sqrt{\sum_{j=1}^n (\frac{w^{(n)}_{j}}{m_n}-\frac{1}{n}  )^2}}\nonumber\\
&=& \frac{F_{m_{n},n}(x) -F_{n}(x)  }{\sqrt{F(x)(1-F(x))} \sqrt{\sum_{j=1}^n (\frac{w^{(n)}_{j}}{m_n}-\frac{1}{n}  )^2}}, \ x\in \mathbb{R}\nonumber
\end{eqnarray}
where
\begin{equation}\label{(5.3)}
F_{m_{n},n}(x):= \sum_{i=1}^n \frac{w^{(n)}_{i}}{m_n} \mathds{1}(X_i \leq x), \ x\in \mathbb{R},
\end{equation}
is the randomized  empirical distribution function.

\par
We note that, point-wise in $x\in \mathds{R}$,
\begin{equation}%\label{(5.4)}
E_{X|w}(F_{m_{n},n}(x))=F(x)= E_{X,w}(F_{m_{n},n}(x)).
\end{equation}

\par
 Define also the randomized sub-sample  variance  of the indicator random variables $\mathds{1}(X_i \leq x)$ by putting
\begin{eqnarray}
S^{{2}}_{m_{n},n}(x)&:=& \sum_{i=1}^{n} w^{(n)}_{i} \big( \mathds{1}(X_i \leq x)- F_{m_{n},n}(x)\big)^2\big/m_n \label{(5.4)}\\
&=& F_{m_{n},n}(x)(1-F_{m_{n},n}(x)), \ x\in \mathbb{R}.\nonumber
\end{eqnarray}

\par
With $n$ \emph{fixed and } $m=m_n\rightarrow +\infty$, along the lines of (\ref{retaning bar{X} with fixed n}) we arrive at

\begin{equation}\label{(5.6)}
F_{m_{n},n}(x)\longrightarrow F_{n}(x)\ in\ probbility-P_{X,w},\ point-wise \ in \ x\in \mathbb{R},
\end{equation}
and, consequently, point-wise in $x \in \mathbb{R}$, \emph{as} $m=m_n\rightarrow +\infty$,
\begin{equation}\label{(5.7)}
S^{{2}}_{m_{n},n}(x) \longrightarrow F_n(x)(1-F_n (x))=S^{2}_{n}(x)\ in \ probability-P_{X,w}.
\end{equation}
Furthermore, \emph{\`{a} la} (\ref{bias of bar{X*}}), \emph{as} $n,m_n\to +\infty$, \emph{point-wise in} $x \in \mathbb{R}$, we conclude

\begin{equation}\label{(5.8)}
\big(F_{m_{n},n}(x)-F_n (x)\big)\longrightarrow 0\ in\ probbility-P_{X,w,}
\end{equation}
that, in turn, \emph{point-wise in} $x\in \mathbb{R}$, \emph{as}  $n,m_n\to +\infty$, implies
\begin{equation}\label{(5.9)}
\big(S^{{2}}_{m_{n},n}-S^{2}_{n}(x) \big) \longrightarrow 0\ in \ probability-P_{X,w},
\end{equation}
with $S^{{2}}_{m_{n},n}$ and $S^{2}_{n}(x)$ respectively as in (\ref{(5.4)}) and (\ref{(5.2)}).

\par
We wish to note and emphasize that, unlike in (\ref{bias of S*}), for concluding (\ref{(5.9)}), \emph{we do not have to assume that} $n=o(m_n)$ as $n,m_n\to +\infty$.
\par

Further to the randomized  standardized  empirical process $\alpha^{(1)}_{n,m_n}(x)$, we now define the following Studentized/self-normalized  versions with $x\in \mathds{R}$, as follows:

\begin{eqnarray}
\hat{\alpha}^{(1)}_{m_{n},n}(x)&:=& \frac{\sum_{i=1}^n (\frac{w^{(n)}_{i}}{m_n}-\frac{1}{n}  ) \mathds{1}(X_i \leq x)}{\sqrt{F_n(x)(1-F_n(x))} \sqrt{\sum_{j=1}^n (\frac{w^{(n)}_{j}}{m_n}-\frac{1}{n}  )^2} } \label{(5.10)}\\
\hat{\hat{\alpha}}^{(1)}_{m_{n},n}(x)&:=& \frac{\sum_{i=1}^n (\frac{w^{(n)}_{i}}{m_n}-\frac{1}{n}  ) \mathds{1}(X_i \leq x)}{\sqrt{F_{m_{n},n}(x)(1-F_{m_{n},n}(x))} \sqrt{\sum_{j=1}^n (\frac{w^{(n)}_{j}}{m_n}-\frac{1}{n}  )^2} } \label{(5.11)} \\
\hat{\alpha}^{(2)}_{m_{n},n}(x)&:=& \frac{\sum_{i=1}^n \big|\frac{w^{(n)}_{i}}{m_n}-\frac{1}{n} \big| \big(\mathds{1}
(X_i \leq x)-F(x)\big)}{\sqrt{F_n(x)(1-F_n(x))} \sqrt{\sum_{j=1}^n (\frac{w^{(n)}_{j}}{m_n}-\frac{1}{n}  )^2} } \label{(5.12)} \\
\hat{\hat{\alpha}}^{(2)}_{m_{n},n}(x)&:=& \frac{\sum_{i=1}^n \big|\frac{w^{(n)}_{i}}{m_n}-\frac{1}{n}  \big| \big(\mathds{1}(X_i \leq x)-F(x)\big)}{\sqrt{F_{m_{n},n}(x)(1-F_{m_{n},n}(x))} \sqrt{\sum_{j=1}^n (\frac{w^{(n)}_{j}}{m_n}-\frac{1}{n}  )^2} }.\label{(5.13)}
\end{eqnarray}
\par
Clearly, on replacing $X_i$ by $\mathds{1}(X_i \leq x)$ and $\mu$ by $F(x)$, $x \in \mathds{R}$,  in the formula  in (\ref{(1.21)}), we arrive at the respective  statements of (\ref{(1.22)new}) and (\ref{(1.22)}) in this context. Also, replacing  $X_i$ by $\mathds{1}(X_i \leq x)$ in the formula as in (\ref{(1.22)'}), we conclude the statement of (\ref{doubleprime}) with $\mu$ replaced by  $F(x)$, $x \in \mathds{R}$.

\par
As to the latter statement, on letting

\begin{equation}\label{(5.15)'}
\hat{F}_{m_{n},n}(x):= \frac{\sum_{i=1}^n |\frac{w^{(n)}_{i}}{m_n}- \frac{1}{n}| \mathds{1}(X_i \leq x) } {\sum_{j=1}^n |\frac{w^{(n)}_{j}}{m_n}- \frac{1}{n}|},
\end{equation}
as $n,m_n \rightarrow +\infty$, such that $m_n=o(n^2)$, point-wise in $x \in \mathds{R}$,  by virtue of (\ref{doubleprime}),

\begin{equation}\label{(5.16)'}
\hat{F}_{m_{n},n}(x) \longrightarrow F(x)\ in \ probability-P_{X,w}.
\end{equation}

\par
In Lemma 5.2 of Cs\"{o}rg\H{o} \emph{et al}. \cite{Scand} it is shown that, \emph{if} $m_n , n\to +\infty$ \emph{so that $m_n=o(n^2)$, then}

\begin{equation}\label{(5.14)}
M_n:=\frac{ \max_{1\leq i \leq n} \big( \frac{w^{(n)}_{i}}{n}-\frac{1}{n}\big)^2 } {\sum_{j=1}^n \big( \frac{w^{(n)}_j}{n}-\frac{1}{n}\big)^2 }\to 0\ in \ probability-P_{w}.
\end{equation}
This, mutatis mutandis, combined with (a) of Corollary 2.1 of Cs\"{o}rg\H{o} \emph{et al}. \cite{Scand}, \emph{as $n, m_n \to +\infty$ so that} $m_n=o(n^2)$, yields
\begin{equation}\label{(5.15)}
P_{X|w} \big( \hat{\alpha}^{(s)}_{m_{n},n}(x)\leq t \big)\to P(Z \leq t) \ in \ probability-P_{w},\ for\ all \ x,t\in \mathbb{R},
\end{equation}
with $s=1$ and also for $s=2$, and via  Lemma 1.2 in  S. Cs\"{o}rg\H{o} and Rosalsky \cite{Csorgo and Rosalsky}, this results in having also the unconditional CLT

\begin{equation}\label{(1.16)}
P_{X,w} \big( \hat{\alpha}^{(s)}_{m_{n},n}(x)\leq t \big)\to P(Z \leq t) \ for\ all \ x,t\in \mathbb{R},
\end{equation}
with $s=1$ and also for $s=2$.

\par
On combining (\ref{(1.16)}) and (\ref{(5.9)}), \emph{as } $n,m_n\to +\infty$ \emph{so that} $m_n=o(n^2)$, when $s=1$ in (\ref{(1.16)}),  we  conclude

\begin{equation}\label{(5.17)}
\hat{\hat{\alpha}}^{(1)}_{m_{n},n}(x)
\substack{d\\ \longrightarrow} Z
\end{equation}
and, when $s=2$ in (\ref{(1.16)}), we arrive at
\begin{equation}\label{(5.18)}
\hat{\hat{\alpha}}^{(2)}_{m_{n},n}(x)
\substack{d\\ \longrightarrow} Z
 \end{equation}
for all $x \in \mathbb{R}$.

\begin{remark}\label{Remark 6.1'}
The Berry-Ess\'{e}en type inequality  (A) of our Theorem \ref{Berry-Esseen} continues to hold true for  $\hat{\alpha}^{(2)}_{m_{n},n}(x)$, and  so does  also (B)   of Theorem \ref{Berry-Esseen} for  $\hat{\alpha}^{(1)}_{m_{n},n}(x)$, without the assumption $E_{X} |X|^{3}<+\infty$, for the indicator random variable $\mathds{1}(X \leq x)$ requires no moments assumptions.
\end{remark}

\begin{remark}
In view of Remark \ref{Remark 6.1'},  in the context of this section, (A) and (B) of Corollary \ref{The Rate} read as follows: As $n,m_n \to +\infty$ in such a way that $m_n=o(n^2)$, then, mutatis mutandis, (A) and (B) hold true for  $\hat{\alpha}^{(1)}_{m_{n},n}(x)$ and ${\hat{\alpha}}^{(2)}_{m_{n},n}(x)$, with $O(\max\{m_n/n^{2}, 1/m_n\})$
in both.  Consequently, statements (\ref{joint dis1}) and (\ref{joint dis2}) of Corollary
\ref{The joint Rate m_n} also read similarly for $\hat{\alpha}^{(1)}_{m_n,n }$ and $\hat{\alpha}^{(2)}_{m_{n},n}(x)$ in terms of the conditions and the rates of convergence.
Thus, on taking $m_n=n$, we immediately obtain the optimal $O(n^{-1})$ rate conclusion of Remark \ref{Remark 2.1} in this context as well, i.e., uniformly in $t \in \mathbb{R}$ and point-wise in $x \in \mathbb{R}$ for $\hat{\alpha}^{(1)}_{m_{n},n}(x)$ and ${\hat{\alpha}}^{(2)}_{m_{n},n}(x)$.
\end{remark}

\begin{remark}\label{6.3'}
As to the  rate of convergence  of the respective CLT's in terms of $P_{X,w}$ as in (\ref{(5.17)}) and (\ref{(5.18)}), and also in terms of $P_{X|w}$,  via (C) and  (D) of Corollary \ref{The Rate}, for  $\hat{\hat{\alpha}}^{(1)}_{m_{n},n}(x)$ and  $\hat{\hat{\alpha}}^{(2)}_{m_{n},n}(x)$,  as $n,m_n \to +\infty$ in such away that $m_n=O(n^2)$,    we obtain the rate $O(\max\{m_n/n^{2}, 1/m_n\})$. Thus,  on taking  $m_n=n$, we conclude the optimal rate of convergence $O(n^{-1})$   for $\hat{\hat{\alpha}}^{(1)}_{m_{n},n}(x)$ and  $\hat{\hat{\alpha}}^{(2)}_{m_{n},n}(x)$,  uniformly in $t \in \mathbb{R}$ and point-wise in $x \in \mathbb{R}$.

\end{remark}

\par
The CLT's for $\hat{\alpha}^{(1)}_{m_n,n}$ and $\hat{\hat{\alpha}}^{(1)}_{m_n,n}$ can be used to construct point-wise confidence sets for the empirical distribution function $F_{n}(.)$, while those for $\hat{\alpha}^{(2)}_{m_n,n}$ and $\hat{\hat{\alpha}}^{(2)}_{m_n,n}$ provide point-wise C.I.'s for the distribution function $F(.)$. We spell out the ones, respectively based on $\hat{\hat{\alpha}}^{(1)}_{m_n,n}$ and $\hat{\hat{\alpha}}^{(2)}_{m_n,n}$, that are valid both in terms of $P_{X|w}$ and $P_{X,w}$ with the rate of convergence $O\big(  \max\{  m_n/n^2, 1/m_n \}  \big)$ (cf. Remark \ref{6.3'}). Thus, as $n,m_n \rightarrow +\infty$ so that $m_n =o(n^2)$, the CLT's in hand respectively result  in  the following asymptotically exact $(1-\alpha)$ size C.I.'s, for any $\alpha \in (0,1)$ and point-wise in $x\in \mathds{R}$:

\begin{equation}\label{(5.22)'}
 F_{m_{n},n}(x)-z_{\alpha/2} S_{m_n, n} (x) \sqrt{.}
\leq F_{n}(x) \leq
 F_{m_{n},n}(x)+z_{\alpha/2} S_{m_n, n} (x) \sqrt{.}
\end{equation}

\begin{equation}\label{(5.23)'}
 \hat{F}_{m_{n},n}(x) - z_{\alpha/2}\frac{S_{m_n, n} (x) \sqrt{.}}{\sum_{j=1}^n |\frac{w_{j}^{(n)}}{m_n}-\frac{1}{n}|}
\leq F(x) \leq
 \hat{F}_{m_{n},n}(x) + z_{\alpha/2}\frac{S_{m_n, n} (x) \sqrt{.}}{\sum_{j=1}^n |\frac{w_{j}^{(n)}}{m_n}-\frac{1}{n}|}
\end{equation}
with $z_{\alpha/2}$ as in (\ref{(1.27)}),  $\sqrt{.}:= \sqrt{\sum_{j=1}^n (\frac{w_{j}^{(n)}}{m_n}-\frac{1}{n})^2}$,  $S_{m_n, n} (x)=F_{m_{n},n}(x)(1-F_{m_{n},n}(x))$ as in (\ref{(5.4)}), $\hat{F}_{m_{n},n}(x)$ as in (\ref{(5.3)}),      and  $F_{m_n,n}(x)$  as in (\ref{(5.15)'}).

\par
On taking $m_n=n$, then, for each $x\in \mathds{R}$, both of the preceding C.I.'s achieve their nominal level at the optimal rate of $O(n^{-1})$. This is a significant achievement in capturing the population distribution by (\ref{(5.23)'}), for each $x\in \mathds{R}$, when the available sample is of moderate size or  small.

\par
In case of having a big data set of size $n$, when processing the entire data set may not be possible,  then both $F_n (.)$ and $F(.)$ are to be estimated. In this case the confidence set (\ref{(5.22)'}) can serve not only for covering  $F_{n}(x)$, but $F(x)$ as well with any desirable accuracy for each $x \in \mathds{R}$.  Namely, on putting  $\varepsilon_n (x)=F_n (x)-F(x)$, $x \in \mathds{R}$, we simply re-write it as follows
\begin{equation}\label{(5.24)'}
F_{m_{n},n}(x)-z_{\alpha/2} S_{m_n, n} (x) \sqrt{.}
\leq F(x)+\varepsilon_n (x) \leq
 F_{m_{n},n}(x)+z_{\alpha/2} S_{m_n, n} (x) \sqrt{.}
\end{equation}
and argue via the Glivenko-Cantelli theorem that in case of big data sets $\varepsilon_n$ is negligible with any desired  accuracy for each $x \in \mathds{R}$ at a fast enough rate of convergence as $n \rightarrow +\infty$, without paying attention to how $m_n$ and $n$  relate to each other when arriving at the asymptotic $(1-\alpha)$ size confidence set that covers $F_n (x)$ for each $x \in \mathds{R}$ as in (\ref{(5.22)'}). This, in turn, is guaranteed by the Dvoretzky-Kiefer-Wolfowitz  \cite{Dvoretzky-Kiefer-Wolfowitz} inequality that asserts for all $\epsilon>0$

\begin{equation}\label{(5.25)'}
P_{X}( \sup_{-\infty < x < +\infty}  | \varepsilon_n (x) |> \epsilon  )\leq 2 \exp(-2 n\epsilon^2).
\end{equation}

\par
On summing in (\ref{(5.25)'}), one concludes the Glivenko-Cantelli theorem at  the indicated exponentially fast rate of convergence to zero  in $P_X$-probability that of course also holds true point-wise in $x\in \mathds{R}$ for $\varepsilon_n (x)$ as in  (\ref{(5.24)'}). Thus, the error induced when estimating $F(x)$, point-wise in $x \in \mathds{R}$, as in (\ref{(5.24)'}) is practically zero for data sets of big size $n$.

\par
For example, in view of inequality (\ref{(5.25)'}),  where the  best possible constant 2 in front of the exponential function is due to Massart \cite{Massart}, when a large sample of size $n=10^6$ is at hand, then   we have 
%for each $x \in \mathds{R}$

\begin{equation}\label{(5.26)'}
P_{X}( \sup_{-\infty < x < +\infty}  |\varepsilon_n (x) |> \epsilon  ) \leq  2 \exp(-2 \epsilon^2 (10^6) )
\end{equation}
for all $\epsilon>0$.  Thus, practically,   the confidence set  (\ref{(5.22)'}) for $F_n (x)$  is also  a C.I. for $F(x)$ in the case of big data sets  of size $n$.

\par
Another spectacular illustration of the negligibility of $\varepsilon_n (x)$ in (\ref{(5.24)'}) is provided by taking $\epsilon=(\log n/n)^{1/2}$ in (\ref{(5.25)'}).

\par
Recall now that as $n,m_n \rightarrow +\infty$ in such  a way that $m_n =o(n^2)$, then the rate of convergence for having the  $(1-\alpha)$ size confidence set (\ref{(5.22)'}) for $F_n (x)$, and also for $F(x)$, in view of (\ref{(5.24)'}), for $x \in \mathds{R}$, is $O\big(   m_n/n^2 , 1/m_n \big)$. Consequently, when drawing a significantly smaller sub-sample of size $m_n=n^{1/2}$, for example, the rate of convergence becomes $O(n^{-1/2})$ that coincides with the rate of convergence of the classical CLT for the Student $t$-statistic and pivot, based on $n$ observations as in (\ref{def. of T_n}) and (\ref{equivalent to T_n}) respectively.  Needless to say that in case of a big data set, a sub-sample of size $m_n =n^{1/2}$   can be a huge reduction in the number of observations that we are to deal with instead of the original sample  that, in our approach,  results in the same magnitude of error as that of the classical CLT when the entire sample  of size $n$ is  to be observed.

\par
To illustrate the reduction provided by our  confidence set  (\ref{(5.22)'}) when it  used to cover $F_n (x)$ or $F(x)$, point-wise in  $x \in \mathds{R}$, we consider a big data set of size $n=10^6$. By generating  the random weights $(w_{1}^{(10^6)},\ldots,w_{10^6}^{(10^6)})$, with $m_{10^6}= \sum_{i=1}^{10^6} w_{i}^{(10^6)}=\sqrt{10^6}=1000$,  independently from the original sample (cf. Remark \ref{Remark 1}), our confidence set (\ref{(5.22)'}) to capture $F_n(x)$   is achieved with   an error proportional to $1/1000$. Recalling also that in this case $\varepsilon_n = F_n (x) -F(x)$ is negligible already (cf. (\ref{(5.26)'})), we also conclude that (\ref{(5.22)'}) captures $F(x)$ with an  error proportional to $1/1000$.

\section{{Proofs}}\label{Proofs}

\subsection*{\normalsize{Proof of Theorem \ref{Berry-Esseen}}}
Due to similarity  of the two cases we only  give  the proof of part (A) of this theorem.
The proof  relies on the fact that, via conditioning on the weights $w^{(n)}_i$'s, $\sum_{i=1}^{n} \big| \frac{w^{(n)}_i}{m_n}- \frac{1}{n}  \big| (X_i-\mu)$ is a sum of independent and non-identically distributed random variables. This in turn enables us to use a Berry-Ess\'{e}en  type inequality for self-normalized sums of independent and non-identically distributed random variables. Also, some of the ideas in the proof are similar to those  of Slutsky's theorem.
\par
We now write
\begin{eqnarray}
G_{m_n,n}^{(1)}&=& \frac{\sum_{i=1}^{n}\big| \frac{w^{(n)}_i}{m_n}-\frac{1}{n}  \big| (X_i -\mu) }{\sigma \sqrt{\sum_{i=1}^{n} (\frac{w^{(n)}_i}{m_n}-\frac{1}{n})^{2} }  } + \frac{\sum_{i=1}^{n}\big| \frac{w^{(n)}_i}{m_n}-\frac{1}{n}  \big| (X_i -\mu) }{\sigma \sqrt{\sum_{i=1}^{n} (\frac{w^{(n)}_i}{m_n}-\frac{1}{n})^{2}}  } \big( \frac{\sigma}{S_n}-1 \big) \nonumber\\
&=:& Z_{m_n} + Y_{m_n}.
\end{eqnarray}
In view of the above setup, for $t\in \mathds{R}$ and $\varepsilon_1>0$, we have
\begin{eqnarray}
-P_{X|w} (|Y_{m_n}|>\varepsilon)&+&P_{X|w}(Z_{m_n} \leq t-\varepsilon)\nonumber\\
&\leq& P_{X|w}(G_{m_n,n}^{(1)} \leq t) \nonumber\\
&\leq & P_{X|w}(Z_{m_n} \leq t+\varepsilon)+ P_{X|w} (|Y_{m_n}|>\varepsilon).\label{eq 1 proofs}
\end{eqnarray}
Observe now that for $\varepsilon_2>0$ we have
\begin{equation}\label{eq 2 proofs}
P_{X|w} (|Y_{m_n}|>\varepsilon)
\leq P_{X|w} \big( |Z_{m_n}| > \frac{\varepsilon}{\varepsilon_1} \big)+ P_{X} \big(|S^{2}_n -\sigma^2|>\varepsilon_{1}^{2} \big).
\end{equation}
One can readily see that
\begin{eqnarray*}
P_{X|w} \big( |Z_{m_n}|> \frac{\varepsilon}{\varepsilon_1} \big) &\leq& (\frac{\varepsilon_2}{\varepsilon_1})^{2} \frac{\sum_{i=1}^{n} (\frac{w^{(n)}_i}{m_n}-\frac{1}{n})^{2} E_{X}(X_1 - \mu)^{2} }{\sigma^2 \sum_{i=1}^{n} (\frac{w^{(n)}_i}{m_n}-\frac{1}{n})^{2} }\\
&=& (\frac{\varepsilon_1}{\varepsilon})^{2}.
\end{eqnarray*}
Combining  now the preceding conclusion with  (\ref{eq 2 proofs}),  (\ref{eq 1 proofs}) can be replaced by
\begin{eqnarray}
&&- (\frac{\varepsilon_1}{\varepsilon})^{2} - P_{X} \big(|S^{2}_n -\sigma^2|>\varepsilon_{1}^{2} \big)+ P_{X|w} (Z_{m_n}\leq t-\varepsilon)\nonumber\\
&\leq& P_{X|w}(G_{m_n,n}^{(1)} \leq t) \nonumber\\
&\leq&(\frac{\varepsilon_1}{\varepsilon})^{2} + P_{X} \big(|S^{2}_n -\sigma^2|>\varepsilon_{1}^{2} \big)+P_{X|w} (Z_{m_n}\leq t+\varepsilon).\label{eq 3 proofs}
\end{eqnarray}
Now, the continuity of the normal distribution $\Phi$ allows us to choose  $\varepsilon_3>0$   so that
$\Phi(t+\varepsilon)-\Phi(t)<\varepsilon_2$ and $\Phi(t-\varepsilon)-\Phi(t)>-\varepsilon_2$. This combined with (\ref{eq 3 proofs}) yields
\begin{eqnarray}
&&- (\frac{\varepsilon_1}{\varepsilon})^{2} - P_{X} \big(|S^{2}_n -\sigma^2|>\varepsilon_{1}^{2} \big)+ P_{X|w} (Z_{m_n}\leq t-\varepsilon)-\Phi(t-\varepsilon)-\varepsilon_2 \nonumber\\
&\leq& P_{X|w}(G_{m_n,n}^{(1)} \leq t)-\Phi(t)\nonumber \\
&\leq&(\frac{\varepsilon_1}{\varepsilon})^{2} + P_{X} \big(|S^{2}_n -\sigma^2|>\varepsilon_{1}^{2} \big)+P_{X|w} (Z_{m_n}\leq t+\varepsilon)-\Phi(t+\varepsilon)+\varepsilon_2.\nonumber\\
\label{eq 4 proofs}
\end{eqnarray}
We now use the Berry-Ess\'{e}en inequality for independent and not necessarily  identically distributed random variables (cf., e.g., Serfling \cite{Serfling}) to write

\begin{equation*}
P_{X|w} (Z_{m_n}\leq t+\varepsilon_1)-\Phi(t+\varepsilon_1)\leq  \frac{  C E_{X}|X-\mu|^3}{\sigma^{3/2}}. \frac{\sum_{i=1}^n  | \frac{w^{(n)}_i}{m_n}-\frac{1}{n}  |^3 }{\big(  \sum_{i=1}^n  ( \frac{w^{(n)}_i}{m_n}-\frac{1}{n}  )^2   \big)^{3/2}}
\end{equation*}
and
\begin{equation*}
P_{X|w} (Z_{m_n}\leq t-\varepsilon_1)-\Phi(t-\varepsilon_1)\geq  \frac{ - C E_{X}|X-\mu|^3}{\sigma^{3/2}}. \frac{\sum_{i=1}^n  | \frac{w^{(n)}_i}{m_n}-\frac{1}{n}  |^3 }{\big(  \sum_{i=1}^n  ( \frac{w^{(n)}_i}{m_n}-\frac{1}{n}  )^2   \big)^{3/2}},
\end{equation*}
where $C$ is a universal constant as in the  Berry-Ess\'{e}en inequality in this context (cf. page 33 of Serfling \cite{Serfling}).
\par
Incorporating these approximations into  (\ref{eq 4 proofs}) we arrive at

\begin{eqnarray*}
&&\sup_{-\infty<t<+\infty} \big| P_{X|w}(G_{m_n,n}^{(1)} \leq t)-\Phi(t)  \big| \nonumber\\
&&\leq  (\frac{\varepsilon_1}{\varepsilon})^{2} + P_{X} \big(|S^{2}_n -\sigma^2|>\varepsilon_{1}^{2} \big)+ \frac{C E_{X}|X-\mu|^3}{\sigma^{3/2}}. \frac{\sum_{i=1}^n  | \frac{w^{(n)}_i}{m_n}-\frac{1}{n}  |^3 }{\big(  \sum_{i=1}^n  ( \frac{w^{(n)}_i}{m_n}-\frac{1}{n}  )^2   \big)^{3/2}}+\varepsilon_2.\nonumber
\end{eqnarray*}

From the preceding  relation  we conclude that

\begin{equation}\label{eq 5 proofs}
P_{w} \big( \sup_{-\infty<t<+\infty} \big| P_{X|w}(G_{m_n,n}^{(1)} \leq t)-\Phi(t)  \big|>\delta    \big) \leq P_{w} \big( \frac{\sum_{i=1}^n  | \frac{w^{(n)}_i}{m_n}-\frac{1}{n}  |^3 }{\big(  \sum_{i=1}^n  ( \frac{w^{(n)}_i}{m_n}-\frac{1}{n}  )^2   \big)^{3/2}} > \delta_n    \big)
\end{equation}
with $\delta_n$  as defined in the statement of  Theorem \ref{Berry-Esseen}.

\par
For $\varepsilon>0$, the right hand side of (\ref{eq 5 proofs}) is  bounded above by

\begin{eqnarray}
 && P_{w}\Big\{  \sum_{i=1}^{n} \big| \frac{w^{(n)}_i}{m_n}-\frac{1}{n}  \big|^3 >\frac{\delta_n(1-\varepsilon)^{\frac{3}{2}}(1-\frac{1}{n})^{\frac{3}{2} } }{m^{\frac{3}{2}}_n} \Big\}\nonumber\\
&&+ P_{w} \big(  \Big| \frac{m_{n}}{1-\frac{1}{n} } \sum_{i=1}^{n} \big( \frac{w^{(n)}_i}{m_n}-\frac{1}{n}  \big)^2 -1  \Big|>\varepsilon   \big)\nonumber\\
&&=: \Pi_1(n)+\Pi_2(n).\nonumber
%&&+ P_{w} \big(  \Big| \frac{m_{n}}{1-\frac{1}{n} } \sum_{i=1}^{n} \big( \frac{w^{(n)}_i}{m_n}-\frac{1}{n}  \big)^2 -1  \Big|>\varepsilon_8   \big)
\end{eqnarray}
We  bound $\Pi_{1}(n)$ above by
\begin{eqnarray}
&&\delta^{-2}_n(1-\varepsilon)^{-3}(1-\frac{1}{n})^{-3 } m^{-3 }_{n}  ( n  + n^2) E_{w} (w^{(n)}_1 -\frac{m_n}{n}    )^6 \nonumber\\
&&= \delta^{-2}_n(1-\varepsilon)^{-3}(1-\frac{1}{n})^{-3 } m^{-3 }_{n}  ( n  + n^2) \{ \frac{15 m^{3}_n}{n^3} + \frac{25 m^{2}_n}{n^2} + \frac{m_n}{n} \}. \label{eq 6 proofs}
\end{eqnarray}

As for $\Pi_{2}(n)$, recalling that $E_{w}\big(\sum_{i=1}^{n} (\frac{w^{(n)}_{i}}{m_n}-\frac{1}{n})^2 \big)=\frac{(1-\frac{1}{n})}{m_n}$,  an application of  Chebyshev's inequality yields
\begin{eqnarray}
\Pi_{2}(n)&\leq& \frac{m^{2}_n}{\varepsilon^{2} (1-\frac{1}{n})^2  } E_{w}\big( \sum_{i=1}^{n} (\frac{w^{(n)}_{i}}{m_n}-\frac{1}{n})^2 - \frac{(1-\frac{1}{n})}{m_n}   \big)^2 \nonumber\\
&=& \frac{m^{2}_n}{\varepsilon^{2} (1-\frac{1}{n})^2  }\ E_{w} \Big\{ \Big(\sum_{i=1}^n (\frac{w^{(n)}_i}{m_n}-\frac{1}{n})^2\Big)^2 - \frac{(1-\frac{1}{n})^2}{m^{2}_n} \Big\}^2  \nonumber\\
&=& \frac{m^{2}_n}{\varepsilon
^{2} (1-\frac{1}{n})^2  } \Big\{ n E_{w} \big( \frac{w^{(n)}_1}{m_n}-\frac{1}{n} \big)^4 + n(n-1) E_{w} \big[\big( \frac{w^{(n)}_1}{m_n}-\frac{1}{n} \big)^2 \big( \frac{w^{(n)}_2}{m_n}-\frac{1}{n} \big)^2\big] \nonumber\\
&-& \frac{(1-\frac{1}{n})^2}{m^{2}_n}   \Big\}.\label{for appendix 2}
\end{eqnarray}
We now use the fact  that $w^{(n)}$'s are multinomially distributed to compute  the preceding relation. After some algebra it turns out that it can be bounded above by
\begin{eqnarray}
&&\frac{m^{2}_n}{\varepsilon^{2} (1-\frac{1}{n})^2  } \Big\{ \frac{1-\frac{1}{n}}{n^3 m^{3}_n } + \frac{(1-\frac{1}{n})^4}{m^{3}_n}  + \frac{(m_n -1)(1-\frac{1}{n})^2}{n m^{3}_n} +  \frac{4(n-1)}{n^3 m_n}       +\frac{1}{m^{2}_n}\nonumber\\
&&~~~~~~~~~~~~~~~~~ - \frac{1}{n m^{2}_n} + \frac{n-1}{n^{3} m^{3}_{n}} + \frac{4(n-1)}{n^2 m^{3}_{n}} - \frac{(1-\frac{1}{n})^2}{m^{2}_n}        \Big\}.\label{eq 7 proofs}
\end{eqnarray}
Incorporating (\ref{eq 6 proofs}) and (\ref{eq 7 proofs}) into (\ref{eq 5 proofs}) completes the proof of
 part (A) of Theorem \ref{Berry-Esseen}. $\square$

\subsection*{\normalsize{Proof of Corollary \ref{The Rate}}}
The proofs of parts (A) and (B) of this corollary are  immediate consequences of Theorem \ref{Berry-Esseen}.
\par
To prove  parts (C) and (D) of this corollary, in view of Theorem \ref{Berry-Esseen} it suffices to show that, for arbitrary  $\varepsilon_1, \varepsilon_2 >0$, as $n,m_n \to +\infty$,
\begin{equation}\label{Edmonton 1}
P_{w}\big( P_{X|w} ( |S_{m_n,n}-S^{2}_{n}|>\varepsilon_1)>\varepsilon_2  \big)=O( \frac{n}{m^{2}_n} ).
\end{equation}
To prove the preceding result we first note that
\begin{eqnarray*}
S^{2}_{m_n,n}-S^{2}_{n}&=&\sum_{1\leq i\neq j \leq n} \big(\frac{w^{(n)}_i w^{(n)}_j}{m_n (m_n -1)}-\frac{1}{n(n-1)} \big) \frac{(X_i-X_j )^2}{2}\\
&=& \sum_{1\leq i\neq j \leq n} \big(\frac{w^{(n)}_i w^{(n)}_j}{m_n (m_n -1)}-\frac{1}{n(n-1)} \big) \big(\frac{(X_i-X_j )^2}{2}-\sigma^2\big).
\end{eqnarray*}
By virtue of the preceding observation, we proceed with the proof of (\ref{Edmonton 1}) by first letting $d^{(n)}_{i,j}:=\frac{w^{(n)}_i w^{(n)}_j}{m_n (m_n -1)}-\frac{1}{n(n-1)}$ and writing

\begin{eqnarray}
&&P_{w}\big\{ P_{X|w} ( \big|  \sum_{1\leq i\neq j \leq n}  d^{(n)}_{i,j} \big(\frac{(X_i-X_j )^2}{2}-\sigma^2\big) \big| >\varepsilon_1)\varepsilon_2  \big\} \nonumber\\
&&\leq P_{w}\big\{ E_{X|w} \big(   \sum_{1\leq i\neq j \leq n}  d^{(n)}_{i,j} \big(\frac{(X_i-X_j )^2}{2}-\sigma^2\big)\big)^2 >\varepsilon^{2}_{1} \varepsilon_2 \big\}.\label{Edmonton 2}
\end{eqnarray}
Observe now that
\begin{eqnarray}
&&E_{X|w} \big(   \sum_{1\leq i\neq j \leq n}  d^{(n)}_{i,j} \big(\frac{(X_i-X_j )^2}{2}-\sigma^2\big)\big)^2 \nonumber \\
&=& E_{X}\big(\frac{(X_1-X_2 )^2}{2}-\sigma^2\big)^2  \sum_{1\leq i\neq j \leq n}  (d^{(n)}_{i,j})^2  \nonumber\\
&+& \sum_{\substack{1\leq i,j,k\leq n\\ i,j,k\ are \ distinct} } d^{(n)}_{i,j} d^{(n)}_{i,k}
 E_{X} \big( (\frac{(X_i-X_j)^2}{2}-\sigma^2)(\frac{(X_i-X_k)^2}{2}-\sigma^2)\big)\nonumber  \\
&+& \sum_{\substack{1\leq i,j,k,l\leq n\\ i,j,k,l\ are\ distinct} } d^{(n)}_{i,j} d^{(n)}_{k,l}
 E_{X} \big( (\frac{(X_i-X_j)^2}{2}-\sigma^2)(\frac{(X_k-X_l)^2}{2}-\sigma^2)\big).\nonumber\\
 \label{Edmonton 3}
\end{eqnarray}
We note that in the preceding relation, since $i,j,k$ are distinct, we have that
\begin{eqnarray*}
&&E_{X} \big( (\frac{(X_i-X_j)^2}{2}-\sigma^2)(\frac{(X_i-X_k)^2}{2}-\sigma^2)\big)\\
&=& E\big\{ E\big( \frac{(X_i-X_j)^2}{2}-\sigma^2|X_i\big) E\big( \frac{(X_i-X_k)^2}{2}-\sigma^2|X_i\big)  \big\}= \frac{E_{X} (X^{2}_{1}-\sigma^2)}{4}.
\end{eqnarray*}
Also, since $i,j,k,l$ are distinct, we have that

\begin{equation*}
E_{X} \big( (\frac{(X_i-X_j)^2}{2}-\sigma^2)(\frac{(X_k-X_l)^2}{2}-\sigma^2) \big)
=  E^{2}_{X} \big( \frac{(X_i-X_j)^2}{2}-\sigma^2)=0.
\end{equation*}
Therefore, in view of (\ref{Edmonton 3}) and (\ref{Edmonton 2}),  the proof of (\ref{Edmonton 1}) follows if we show that

\begin{equation}\label{Edmonton 4}
\sum_{1\leq i\neq j \leq n}  (d^{(n)}_{i,j})^2=O_{P_{w}}( \frac{1}{m^{2}_n})
\end{equation}
and
\begin{equation}\label{Edmonton 5}
\sum_{\substack{1\leq i,j,k\leq n\\ i,j,k\ are \ distinct} } d^{(n)}_{i,j} d^{(n)}_{i,k}=O_{P_{w}}(\frac{n}{m^{2}_n}).
\end{equation}
Noting that, as $n,m_n\to +\infty$,
\begin{equation*}
E_{w}\big\{ \sum_{1\leq i\neq j \leq n} (d^{(n)}_{i,j})^2\big\} \sim \frac{1}{m^{2}_n}
\end{equation*}
and

\begin{equation*}
E_{w} \big|  \sum_{\substack{1\leq i,j,k\leq n\\ i,j,k\ are \ distinct} } d^{(n)}_{i,j} d^{(n)}_{i,k} \big| \leq n^3 E_{w}(d^{(n)}_{1,2})^2 \sim \frac{n}{m^{2}_{n}}.
\end{equation*}
The  preceding two conclusions imply (\ref{Edmonton 4}) and (\ref{Edmonton 5}), respectively.
Now the proof of Corollary \ref{The Rate} is complete. $\square$

\subsection*{Proof of Corollary \ref{The joint Rate m_n}}
The proof of this result is relatively easy. Due to their similarity, we only give the proof  for part (A) as follows.    For arbitrary positive $\delta$, we write

\begin{eqnarray*}
\sup_{-\infty<t<+\infty} \big|  P_{X,w}\big( G_{m_n,n}^{(1)}\leq t \big)-\Phi(t)   \big| &\leq& \delta+ 2 P\big( \big| P_{X|w}(G_{m_n,n}^{(1)}\leq t) -\Phi(t)    \big|>\delta  \big)\\
&=& \delta+ O\Big( \max\{ \frac{m_n}{n^2}, \frac{1}{m_n} \}  \Big), \ as \ n,m_n \rightarrow +\infty.
\end{eqnarray*}
The last relation above is true in view of Corollary \ref{The Rate}. $\square$

\section{\normalsize{Appendix 1}}
Consider the original sample $\{X_1,\ldots,X_n\}$ and assume that the sample size $n\geq 1$ is fixed.
We are now to show that when $n$ is fixed, as $m\rightarrow +\infty$, we have $\hat{X}_{m_n,n}\rightarrow \bar{X}_{n}$ in probability $P_{X,w}$. To do so, without loss of generality we assume that $\mu=0$. Let $\varepsilon_1,\varepsilon_2>0$,  and write
\begin{eqnarray}
P_{w} \big\{ P_{X|w} \big( \big|\hat{X}_{m}- \bar{X}_{n}\big|> \varepsilon_1 \big)>\varepsilon_2   \big\}
&\leq&    P_{w} \big\{ E_{X|w} \big( \sum_{i=1}^n (\frac{w^{(n)}_i}{m}-\frac{1}{n}) X_i \big)^2 >\varepsilon^{2}_1 \varepsilon_2     \big\}\nonumber\\
&=& P_{w} \big\{ \sum_{i=1}^n  (\frac{w^{(n)}_i}{m}-\frac{1}{n}) ^2 >  \sigma^{-2} \varepsilon^{2}_1 \varepsilon_2   \big\}\nonumber\\
&\leq& \sigma^{2} \varepsilon^{-2}_1 \varepsilon^{-1}_2 \ n E_{w} \big( \frac{w^{(n)}_1}{m}-\frac{1}{n} \big)^{2}\nonumber\\
&\leq& \sigma^{-2} \varepsilon^{-2}_1 \varepsilon^{-1}_2 \frac{(1-\frac{1}{n})}{m}\rightarrow 0, \ as\ m\rightarrow \infty. \nonumber\\
\label{(5.30)}
\end{eqnarray}
The preceding conclusion means that $P_{X|w} \big( \big|\hat{X}_{m_n}- \bar{X}_{n}\big|> \varepsilon_1 \big)\rightarrow 0$ in probability-$P_{w}$. Hence, by  the dominated convergence theorem,  we conclude that $\hat{X}_{m}\rightarrow \bar{X}_{n}$ in probability $P_{X,w}$.  $\square$
\par
We are now to show that the randomized  sample variance  $S^{2}_{m_n,n}$ is an  in probability consistent estimator of the ordinary sample variance $S^{2}_{n}$ for each fixed $n$, when $m\rightarrow+\infty$.
Employing now the $u$-statistic representation of the sample variance  enables us to rewrite $S^{2}_{m_n,n}$, as in (\ref{def. of S^*}), as follows
\begin{eqnarray*}
S^{2}_{m_n,n}&=& \frac{\sum_{1\leq i\leq j\leq n } w^{(n)}_i w^{(n)}_j (X_{i}-X_{j})^{2} }{2 m(m-1)}.
\end{eqnarray*}
In view of the preceding formula, we have
\begin{equation}\nonumber
 S^{2}_{m_n,n}-S^{2}_n=\sum_{1\leq i\neq j\leq n} \big( \frac{1}{2m(m-1)}-\frac{1}{2n(n-1)}\big) (X_i-X_j)^2.
\end{equation}
Now, for $\ep_1,\ep_2>0$,  we write

\begin{eqnarray}
&&P_{w}\Big( P_{X|w}\big(\Big| S^{2}_{m_n,n}-S^{2}_n  \Big|>\ep_1\big)>\ep_2 \Big)  \nonumber\\
&&=P_w\big(  P_{X|w}\big( \big|\mathop{\sum\sum}_{1\leq i\neq j\leq n}
(\frac{w_i^{(n)}w_j^{(n)}}{m(m-1)}
- \frac{1}{ n(n-1)} ) (X_i-X_j)^2\big| > 2 \ep_1\big)>\ep_2  \big)\nonumber \\
&\leq&  P_w\big(   \mathop{\sum\sum}_{1\leq i\neq j\leq n}
\big| \frac{w_i^{(n)}w_j^{(n)}}{m_n(m_n-1)}
- \frac{1}{ n(n-1)} \big| E_{X}(X_i-X_j)^2 > 2 \ep_1 \ep_2  \big)\nonumber \\
&\leq&  P_w\big(   \mathop{\sum\sum}_{1\leq i\neq j\leq n}
\big| \frac{w_i^{(n)}w_j^{(n)}}{m_n(m_n-1)}
- \frac{1}{ n(n-1)} \big|  > \ep_1 \ep_2 \sigma^{-2}  \big).\label{(5.31)}
\end{eqnarray}

The preceding relation can be bounded above by:
\begin{eqnarray}
&& \ep^{-2}_1 \ep^{-2}_2 \sigma^{4}\Big\{ n(n-1) E_{w} \big(  \frac{w^{(n)}_1 w^{(n)}_2}{m (m-1)}-\frac{1}{n(n-1)} \big)^2 \nonumber \\
&+& n(n-1)(n-2) E_{w}\Big( \big|\frac{w^{(n)}_1 w^{(n)}_2}{m (m-1)}-\frac{1}{n(n-1)}\big|   \big|\frac{w^{(n)}_1 w^{(n)}_3}{m (m-1)}-\frac{1}{n(n-1)}\big|   \Big) \nonumber\\
&+& n(n-1)(n-2)(n-3) E_{w}\Big( \big|\frac{w^{(n)}_1 w^{(n)}_2}{m (m-1)}-\frac{1}{n(n-1)}\big|   \big|\frac{w^{(n)}_3 w^{(n)}_4}{m (m-1)}-\frac{1}{n(n-1)}\big|   \Big)  \Big\} \nonumber
\\
&\leq& \ep^{-2}_1 \ep^{-2}_2 \sigma^{4} \Big\{ n(n-1) E_{w} \big(  \frac{w^{(n)}_1 w^{(n)}_2}{m (m-1)}-\frac{1}{n(n-1)} \big)^2 \nonumber\\
&+& n(n-1)(n-2) E_{w} \big(  \frac{w^{(n)}_1 w^{(n)}_2}{m (m-1)}-\frac{1}{n(n-1)} \big)^2 \nonumber \\
&+& n(n-1)(n-2)(n-3)  E_{w} \big(  \frac{w^{(n)}_1 w^{(n)}_2}{m(m -1) } -\frac{1}{n(n-1)} \big)^2 \Big\} \nonumber \\
&=& \ep^{-2}_1 \ep^{-2}_2 \sigma^{4}\big\{ n(n-1)\nonumber\\
&+& n(n-1)(n-2)+ n(n-1)(n-2)(n-3) \big\}  \Big\{ \frac{1}{n^4 m^{2}  } + \frac{n}{n^4 m^{2} } + \frac{n^2}{ n^4 m^{2}} \Big\}.\nonumber
 \end{eqnarray}
Clearly, the latter  term approaches zero when $m\rightarrow +\infty$, for each fixed $n$. By this we have shown that $S^{2}_{m_n,n}\rightarrow S^{2}_{n}$ in probability-$P_{X,w}$, when $n$ is fixed and only $m\rightarrow +\infty$. $\square$

\subsection*{\normalsize{Consistency of $ \hat{X}_{m_n,n}$ in (\ref{(1.22)'})} }
We give the proof of  (\ref{(1.22)'}) for $m_n=n$, noting that the proof  below  remains the same for $m_n\leq n$ and it can  be  adjusted for the case $m_n= k n$, where $k$ is a positive integer.
In order to  establish (\ref{(1.22)'}) when $m_n=n$, we first note that
\begin{equation*}
E_{X|w}(\sum_{i=1}^n|\frac{w^{(n)}_i}{n}-\frac{1}{n}|)=2(1-\frac{1}{n})^{n}
\end{equation*}
and, with  $\varepsilon_1, \varepsilon_2, \varepsilon_3>0$, we proceed  as follows.
\begin{eqnarray*}
&&P_{w} \big\{ P_{X|w} \big( \big| \hat{X}_{m_n,n}-\mu   \big|>\varepsilon_1        \big)>\varepsilon_2   \big\}\\
&\leq& P_{w}\big\{ P_{X|w} \Big( \big| \hat{X}_{m_n,n}-\mu   \big|>\varepsilon_1        \Big)>\varepsilon_2,  \big| \sum_{=1}^n|\frac{w^{(n)}_j}{n}-\frac{1}{n}|-2(1-\frac{1}{n})^{n}\big|\leq \varepsilon_3   \big\}\\
&+& P_{w} \big\{  \big| \sum_{j=1}^n|\frac{w^{(n)}_j}{n}-\frac{1}{n}|-2(1-\frac{1}{n})^{n}\big|>\varepsilon_3    \big\}\\
&\leq& P_{w} \big\{ P_{X|w} \big( \big|\sum_{i=1}^n |\frac{w^{(n)}_j}{n}-\frac{1}{n}|(X_i-\mu) \big|>  \varepsilon_1 \big(2(1-\frac{1}{n})^{n}-\varepsilon_3\big)          \big)>\varepsilon_2 \big\}\\
&+& \varepsilon_{3}^{-2}  E_{w} \big( \sum_{j=1}^n|\frac{w^{(n)}_j}{n}-\frac{1}{n}|-2(1-\frac{1}{n})^{n}   \big)^2\\
&\leq& P_{w} \big\{ \sum_{i=1}^n  (\frac{w^{(n)}_i}{n}-\frac{1}{n}) ^2 >  \sigma^{-2} \big(2(1-\frac{1}{n})^{n}-\varepsilon_3\big)^{2} \varepsilon_2   \big\}\\
&+& \varepsilon_{3}^{-2} \big\{  n E_{w} (\frac{w^{(n)}_1}{n}-\frac{1}{n}) ^2 +n(n-1) E_{w}\big( \big|  \frac{w^{(n)}_1}{n}-\frac{1}{n} \big|  \big| \frac{w^{(n)}_2}{n}-\frac{1}{n}  \big|  \big) - 4  (1-\frac{1}{n})^{2n}   \big\}\\
&=:& K_1(n)+K_2(n).
\end{eqnarray*}
A similar argument to that in (\ref{(5.30)}) implies that, as $n\to +\infty$, and then $\varepsilon_3\to 0$, we have  $K_1(n) \to 0$. As to  $K_2 (n)$, we note that
\begin{eqnarray*}
E_{w} (\frac{w^{(n)}_1}{n}-\frac{1}{n}) ^2&=&n^{-2} (1-\frac{1}{n})\\
E_{w}\big( \big|  \frac{w^{(n)}_1}{n}-\frac{1}{n} \big|  \big| \frac{w^{(n)}_2}{n}-\frac{1}{n}  \big|  \big)&=& -n^{-3}+4 n^{-2} (1-\frac{1}{n})^n (1-\frac{1}{n-1})^n.
\end{eqnarray*}
Observing now that, as $n\to +\infty$,
\begin{equation*}
n(n-1)E_{w}\big( \big|  \frac{w^{(n)}_1}{n}-\frac{1}{n} \big|  \big| \frac{w^{(n)}_2}{n}-\frac{1}{n}  \big|  \big)-4 (1-\frac{1}{n})^{2n}\to 0,
\end{equation*}
we conclude  that, as $n \to +\infty$, $K_2(n)\to 0$. By this we have concluded the consistency of  $\hat{X}_{m_n,n}$ for the population mean $\mu$, when $m_n=n$. $\square$

\section{\normalsize{Appendix 2}}
The convergence in distribution of the partial sums of the form $\sum_{i=1}^n w_{i}^{(n)} X_{i}$ associated with $T^{(i)}_{m_n,n}$, $i=1,2$,    were also studied in the context of the bootstrap by Cs\"{o}rg\H{o} \emph{et al.} \cite{Scand} via conditioning on the weights  (cf. Theorem 2.1 and Corollary 2.2 therein).
We note that  the latter results include only randomly weighted statistics that are similar to $T_{m_n,n}^{(i)}$, $i=1,2$,  which are natural pivots for the sample mean  $\bar{X}_n$. In view of the fact that $G^{(i)}_{m_n,n}$, $i=1,2$, as defined by (\ref{G^{*}}) and (\ref{G^{**}}), are  natural pivots for the population mean $\mu:=E_X X$, in a similar fashion to Theorem 2.1 and its Corollary 2.2 of  Cs\"{o}rg\H{o} \emph{et al.} \cite{Scand}, here we state    conditional CLT's, given the weights $w_{i}^{(n)}$'s, where $(w_{1}^{(n)},\ldots, w_{i}^{(n)})\ {\substack{d\\=}\ multinomial (m_n; \ 1/n,\ldots,1/n)}$, for the partial sums $\sum_{i=1}^n |\frac{w^{(n)}_{i}}{m_n} - \frac{1}{n}|(X_i-\mu)$. The proofs of these results are essentially identical to that of  Corollary 2.2 of  Cs\"{o}rg\H{o} \emph{et al.} \cite{Scand} in view of the more general setup in terms of  notations in the latter paper.

\begin{thm}\label{thm1}
Let $X,X_1,\ldots$ be real valued i.i.d. random variables with mean $\mu$ and variance $\sigma^2$, where  $0< \sigma^{2} <+\infty$.
\\
(a) If  $m_{n}, n \rightarrow \infty$, in such a way that $m_n=o(n^2)$, then
\begin{equation*}
P_{X|w} (G^{(1)}_{m_n,n} \leq t) \longrightarrow
  \Phi(t)\ in \ probability-P_w \ for\ all\ t \in \mathds{R}.
\end{equation*}
(b) If  $m_{n}, n \rightarrow \infty$ in such a way that $m_n=o(n^2)$ and $ n=o(m_n)$, then
\begin{equation*}
 P_{X|w} (G^{(2)}_{m_n,n} \leq t) \longrightarrow
  \Phi(t)\ in \ probability-P_w, for\ all\ t \in  \mathds{R}.
 \end{equation*}
\end{thm}

%  \slink[url]{http://lib.stat.cmu.edu/aoas/???/???}
%  \sdescription{Some text}

%\begin{supplement}[id=suppA]
%  \sname{Supplement B}
%  \stitle{Appendix 2}
%\end{supplement}
%By using the The rate of Corollary \ref{The Rate} of
\end{document}